\documentclass[a4paper,11pt]{article}
\usepackage{a4p}
\usepackage{pennames}
\usepackage{citesort}
\usepackage[final]{graphics}

\renewcommand{\arraystretch}{1.1}

\newcommand{\GeV}{\ensuremath{\mathrm{GeV}}}
\newcommand{\TeV}{\ensuremath{\mathrm{TeV}}}
\newcommand{\fs}{\ensuremath{\mathrm{fs}}}
\newcommand{\pb}{\ensuremath{\mathrm{pb}}}

\newcommand{\gfermi}{\ensuremath{G_\mathrm{F}}}
\newcommand{\alphas}{\ensuremath{\alpha_\mathrm{s}}}

\newcommand{\mc}{\multicolumn{1}{c}}

\newcommand{\nn}{\nonumber}
\newcommand{\dd}{\mathrm{d}}
\newcommand{\m}{\ensuremath{\phantom{-}}}

\newcommand{\ra}{\ensuremath{\rightarrow}}
\newcommand{\ot}{\ensuremath{\otimes}}

\newcommand{\Pgtm}{\ensuremath{\mathrm{\tau^-}}}
\newcommand{\Pgtp}{\ensuremath{\mathrm{\tau^+}}}

\newcommand{\Pt}{\ensuremath{\mathrm{t}}}

\newcommand{\Pl}{\ensuremath{\mathrm{\ell}}}
\newcommand{\Pal}{\ensuremath{\mathrm{\overline{\ell}}}}
\newcommand{\Pgnl}{\ensuremath{\mathrm{\nu}_{\ell}}}

\newcommand{\Ph}{\ensuremath{\mathrm{h}}}

\newcommand{\Ptau}{\ensuremath{P_\Pgt}}

\renewcommand{\Pgr}{\ensuremath{\mathrm{\rho}}}
\renewcommand{\Pai}{\ensuremath{\mathrm{a_1}}}

\newcommand{\PWL}{\ensuremath{\mathrm{W_L}}}
\newcommand{\PWi}{\ensuremath{\mathrm{W_1}}}
\newcommand{\PWii}{\ensuremath{\mathrm{W_2}}}

\newcommand{\tautol}{\ensuremath{\Pgt\ra\Pl\,\Pgnl\Pgngt}}
\newcommand{\tautoe}{\ensuremath{\Pgt\ra\Pe\,\Pgne\Pgngt}}
\newcommand{\tautom}{\ensuremath{\Pgt\ra\Pgm\,\Pgngm\Pgngt}}
\newcommand{\tautoh}{\ensuremath{\Pgt\ra\Ph\,\Pgngt}}
\newcommand{\tautop}{\ensuremath{\Pgt\ra\Pgp\,\Pgngt}}
\newcommand{\tautor}{\ensuremath{\Pgt\ra\Pgr\,\Pgngt}}
\newcommand{\tautoa}{\ensuremath{\Pgt\ra\Pai\,\Pgngt}}

\newcommand{\eeee}{\ensuremath{\Pep\Pem\ra\Pep\Pem}}
\newcommand{\eemm}{\ensuremath{\Pep\Pem\ra\Pgmp\Pgmm}}
\newcommand{\eett}{\ensuremath{\Pep\Pem\ra\Pgtp\Pgtm}}
\newcommand{\EEEE}{\ensuremath{\Pe\,\Pe\ra\Pe\,\Pe}}
\newcommand{\EEMM}{\ensuremath{\Pe\,\Pe\ra\Pgm\Pgm}}

\newcommand{\ggee}{\ensuremath{\Pgg\Pgg\ra\Pep\Pem}}
\newcommand{\ggmm}{\ensuremath{\Pgg\Pgg\ra\Pgmp\Pgmm}}

\newcommand{\ggEE}{\ensuremath{\Pgg\Pgg\ra\Pe\,\Pe}}
\newcommand{\ggMM}{\ensuremath{\Pgg\Pgg\ra\Pgm\Pgm}}

\newcommand{\tauitoe}[1]{\ensuremath{\Pgt_{#1}\ra\Pe\,\Pgne\Pgngt}}
\newcommand{\tauitom}[1]{\ensuremath{\Pgt_{#1}\ra\Pgm\,\Pgngm\Pgngt}}
\newcommand{\tauitoh}[1]{\ensuremath{\Pgt_{#1}\ra\Ph\,\Pgngt}}

\newcommand{\gvlr }{\ensuremath{g_\mathrm{LR}^\mathrm{V }}}
\newcommand{\gvrl }{\ensuremath{g_\mathrm{RL}^\mathrm{V }}}
\newcommand{\gvll }{\ensuremath{g_\mathrm{LL}^\mathrm{V }}}
\newcommand{\gvrr }{\ensuremath{g_\mathrm{RR}^\mathrm{V }}}

\newcommand{\gslr }{\ensuremath{g_\mathrm{LR}^\mathrm{S }}}
\newcommand{\gsrl }{\ensuremath{g_\mathrm{RL}^\mathrm{S }}}
\newcommand{\gsll }{\ensuremath{g_\mathrm{LL}^\mathrm{S }}}
\newcommand{\gsrr }{\ensuremath{g_\mathrm{RR}^\mathrm{S }}}
\newcommand{\gslrs}{\ensuremath{g_\mathrm{LR}^\mathrm{S*}}}
\newcommand{\gsrls}{\ensuremath{g_\mathrm{RL}^\mathrm{S*}}}
\newcommand{\gslls}{\ensuremath{g_\mathrm{LL}^\mathrm{S*}}}
\newcommand{\gsrrs}{\ensuremath{g_\mathrm{RR}^\mathrm{S*}}}

\newcommand{\gtlr }{\ensuremath{g_\mathrm{LR}^\mathrm{T }}}
\newcommand{\gtrl }{\ensuremath{g_\mathrm{RL}^\mathrm{T }}}
\newcommand{\gtlrs}{\ensuremath{g_\mathrm{LR}^\mathrm{T*}}}
\newcommand{\gtrls}{\ensuremath{g_\mathrm{RL}^\mathrm{T*}}}

\newcommand{\gsrrl}{\ensuremath{g_\mathrm{RR,\ell}^\mathrm{S}}}

\newcommand{\gs   }{\ensuremath{g^\mathrm{S}}}
\newcommand{\gv   }{\ensuremath{g^\mathrm{V}}}
\newcommand{\gt   }{\ensuremath{g^\mathrm{T}}}

\newcommand{\gsmax}{\ensuremath{g^\mathrm{S}_\mathrm{max}}}
\newcommand{\gvmax}{\ensuremath{g^\mathrm{V}_\mathrm{max}}}
\newcommand{\gtmax}{\ensuremath{g^\mathrm{T}_\mathrm{max}}}

\newcommand{\xd}    {\ensuremath{\xi\delta}}

\newcommand{\rhou}  {{\ensuremath{\rho}}}
\newcommand{\xiu}   {{\ensuremath{\xi}}}
\newcommand{\xdu}   {{\ensuremath{\xi\delta}}}
\newcommand{\etau}  {{\ensuremath{\eta}}}

\newcommand{\rhol}  {{\ensuremath{\rho_\Pl}}}
\newcommand{\xil}   {{\ensuremath{\xi_\Pl}}}
\newcommand{\xdl}   {{\ensuremath{(\xi\delta)_\Pl}}}
\newcommand{\etal}  {{\ensuremath{\eta_\Pl}}}

\newcommand{\rhoe}  {{\ensuremath{\rho_\Pe}}}
\newcommand{\xie}   {{\ensuremath{\xi_\Pe}}}
\newcommand{\xde}   {{\ensuremath{(\xi\delta)_\Pe}}}
\newcommand{\etae}  {{\ensuremath{\eta_\Pe}}}

\newcommand{\rhom}  {{\ensuremath{\rho_\Pgm}}}
\newcommand{\xim}   {{\ensuremath{\xi_\Pgm}}}
\newcommand{\xdm}   {{\ensuremath{(\xi\delta)_\Pgm}}}
\newcommand{\etam}  {{\ensuremath{\eta_\Pgm}}}

\newcommand{\deltal}{{\ensuremath{\delta_\Pl}}}
\newcommand{\deltae}{{\ensuremath{\delta_\Pe}}}

\newcommand{\Ntaupairs}{$147\,042$} 

\newcommand\Neh{$  19369 $}
\newcommand\EFFeh{$  83.1\,\% $}
\newcommand\BKGeh{$   3.3\,\% $}
\newcommand\BKGehee{$  0.73\,\% $}
\newcommand\BKGehem{$  1.70\,\% $}

\newcommand\BKGehEE{$  0.06\,\% $}

\newcommand\BKGehEEEE{$  0.03\,\% $}

\newcommand\BKGehhh{$  0.72\,\% $}

\newcommand\Nmh{$  21190 $}
\newcommand\EFFmh{$  88.6\,\% $}
\newcommand\BKGmh{$   5.0\,\% $}

\newcommand\BKGmhme{$  0.99\,\% $}
\newcommand\BKGmhmm{$  1.64\,\% $}

\newcommand\BKGmhMM{$  0.56\,\% $}

\newcommand\BKGmhEEMM{$  0.21\,\% $}

\newcommand\BKGmhhh{$  1.55\,\% $}

\newcommand\Nem{$   5834 $}
\newcommand\EFFem{$  85.8\,\% $}
\newcommand\BKGem{$   2.8\,\% $}
\newcommand\BKGemee{$  0.00\,\% $}
\newcommand\BKGemmm{$  0.12\,\% $}

\newcommand\BKGemEE{$  0.00\,\% $}
\newcommand\BKGemMM{$  0.06\,\% $}
\newcommand\BKGemEEEE{$  0.00\,\% $}
\newcommand\BKGemEEMM{$  0.04\,\% $}

\newcommand\BKGemeh{$  1.73\,\% $}
\newcommand\BKGemhm{$  0.78\,\% $}

\newcommand\RHOE{$  0.779 \pm  0.047 $}
\newcommand\XIE {$   1.13 \pm   0.39 $}
\newcommand\XDE {$   0.72 \pm   0.31 $}

\newcommand\RHOM{$  0.777 \pm  0.044 $}
\newcommand\XIM {$   0.79 \pm   0.41 $}
\newcommand\XDM {$   0.63 \pm   0.23 $}
\newcommand\ETAM{$  0.010 \pm  0.065 $}

\newcommand\RHOEvXIE {$ -0.716 $}
\newcommand\RHOEvXDE {$ -0.700 $}

\newcommand\RHOEvRHOM{$ -0.249 $}
\newcommand\RHOEvXIM {$  0.229 $}
\newcommand\RHOEvXDM {$  0.344 $}
\newcommand\RHOEvETAM{$ -0.035 $}

\newcommand\XIEvXDE  {$  0.592 $}

\newcommand\XIEvRHOM {$  0.251 $}
\newcommand\XIEvXIM  {$ -0.337 $}
\newcommand\XIEvXDM  {$ -0.307 $}
\newcommand\XIEvETAM {$ -0.006 $}

\newcommand\XDEvRHOM {$  0.321 $}
\newcommand\XDEvXIM  {$ -0.289 $}
\newcommand\XDEvXDM  {$ -0.442 $}
\newcommand\XDEvETAM {$  0.037 $}

\newcommand\RHOMvXIM {$ -0.427 $}
\newcommand\RHOMvXDM {$ -0.590 $}
\newcommand\RHOMvETAM{$  0.461 $}

\newcommand\XIMvXDM  {$  0.271 $}
\newcommand\XIMvETAM {$  0.132 $}

\newcommand\XDMvETAM {$ -0.116 $}

\newcommand\RHOU{$  0.781 \pm  0.028 $}
\newcommand\XIU {$   0.98 \pm   0.22 $}
\newcommand\XDU {$   0.65 \pm   0.14 $}
\newcommand\ETAU{$  0.027 \pm  0.055 $}

\newcommand\RHOUvXIU {$ -0.521 $}
\newcommand\RHOUvXDU {$ -0.492 $}
\newcommand\RHOUvETAU{$  0.348 $}

\newcommand\XIUvXDU  {$  0.273 $}
\newcommand\XIUvETAU {$ -0.022 $}

\newcommand\XDUvETAU {$ -0.094 $}

\newcommand\LIMgsrrE{$   1.36 $}
\newcommand\LIMgslrE{$   1.40 $}
\newcommand\LIMgsrlE{$   2.00 $}

\newcommand\LIMgvrrE{$   0.68 $}
\newcommand\LIMgvlrE{$   0.43 $}
\newcommand\LIMgvrlE{$   0.56 $}

\newcommand\LIMgtlrE{$   0.41 $}
\newcommand\LIMgtrlE{$   0.52 $}

\newcommand\LIMgsrrM{$   1.25 $}
\newcommand\LIMgslrM{$   1.27 $}
\newcommand\LIMgsrlM{$   2.00 $}

\newcommand\LIMgvrrM{$   0.62 $}
\newcommand\LIMgvlrM{$   0.39 $}
\newcommand\LIMgvrlM{$   0.55 $}

\newcommand\LIMgtlrM{$   0.37 $}
\newcommand\LIMgtrlM{$   0.52 $}

\newcommand\LIMgsrrU{$   1.05 $}
\newcommand\LIMgslrU{$   1.10 $}
\newcommand\LIMgsrlU{$   2.00 $}

\newcommand\LIMgvrrU{$   0.53 $}
\newcommand\LIMgvlrU{$   0.35 $}
\newcommand\LIMgvrlU{$   0.52 $}

\newcommand\LIMgtlrU{$   0.32 $}
\newcommand\LIMgtrlU{$   0.51 $}

\newcommand\QtauRU{$  0.089 \pm  0.131 <  0.304$\quad (90\,\% C.L.) }

\newcommand{\LIMHiggs}{$ m_{H^\pm} > 0.97\,\times\,\tan\beta $ \,GeV\quad (95\,\% C.L.)}

\newcommand{\LIMWR}{\ensuremath{m_{\PWR} > 145\,\GeV}\quad(95\,\% C.L.)}
\newcommand{\LIMWii}{\ensuremath{m_{\PWii} > 137\,\GeV}\quad(95\,\% C.L.)}
\newcommand{\LIMzeta}{\ensuremath{|\zeta| < 0.12}\quad(95\,\% C.L.)}

\newcommand\SRHOE  {0.029}
\newcommand\SXIE   {0.14}
\newcommand\SDELTAE{0.14}

\newcommand\SRHOM  {0.016}
\newcommand\SXIM   {0.05}
\newcommand\SDELTAM{0.09}
\newcommand\SETAM  {0.001}

\newcommand\SRHOU  {0.018}
\newcommand\SXIU   {0.10}
\newcommand\SDELTAU{0.07}
\newcommand\SETAU  {0.005}

\newcommand{\TAUOLA}     {\textsc{tauola}}
\newcommand{\KORALZ}     {\textsc{koralz}}
\newcommand{\BABAMC}     {\textsc{babamc}}
\newcommand{\PHOTOS}     {\textsc{photos}}
\newcommand{\ZFITTER}    {\textsc{zfitter}}
\newcommand{\GEANT}      {\textsc{geant}}
\newcommand{\MINUIT}     {\textsc{minuit}}
\newcommand{\VERMASEREN} {\textsc{vermaseren}}

\newcommand{\Lumi}{\ensuremath{155\,\pb^{-1}}}

\newcommand{\BRAmu}{\ensuremath{0.1736 \pm 0.0027}}
\newcommand{\TAUtau}{\ensuremath{(289.2 \pm 2.1)\,\fs}}

\newcommand{\MW}{\ensuremath{(80.43 \pm 0.08)\,\GeV}}
\newcommand{\MZ}{\ensuremath{(91.1884 \pm 0.0022)\,\GeV}}
\newcommand{\Mt}{\ensuremath{(175 \pm 6)\,\GeV}}

\newcommand{\Alphas}{\ensuremath{0.118 \pm 0.003}}
\newcommand{\Alpha}{\ensuremath{128.90 \pm 0.09}}

\begin{document}

\begin{titlepage}
\begin{center}{\large   EUROPEAN LABORATORY FOR PARTICLE PHYSICS
}\end{center}\bigskip
\begin{flushright}
       CERN-EP/98-104 \\ June 29, 1998
\end{flushright}
\bigskip\bigskip\bigskip\bigskip\bigskip
\begin{center}{\huge\bf
  Measurement of the Michel Parameters \\ in Leptonic Tau Decays
  \rule{0em}{2.5ex}
}\end{center}\bigskip\bigskip
\begin{center}{\LARGE The OPAL Collaboration
}\end{center}\bigskip\bigskip
\bigskip\begin{center}{\large  Abstract}\end{center}
\noindent
The Michel parameters of the leptonic \Pgt\ decays are measured using
the OPAL detector at LEP\@. The parameters \rhol, \xil, \xdl\ (with
$\Pl = \Pe,\Pgm$) and \etam\ are extracted from the energy spectra of
the charged decay leptons and from their energy-energy correlations. A
new method involving a global likelihood fit of Monte Carlo generated
events with complete detector simulation and background treatment has
been applied to the data recorded at center-of-mass energies close to
$\sqrt{s}=m_\PZz$ corresponding to an integrated luminosity of \Lumi.
If \Pe-\Pgm\ universality is assumed and inferring the \Pgt\ 
polarization from neutral current data, the measured Michel parameters
are:
\begin{center}
  \begin{tabular}{r@{$\ \ =\ \ $}l@{\qquad}r@{$\ \ =\ \ $}l}
    \rhou & $\mbox{\RHOU} \pm \SRHOU   $, &
    \xiu  & $\mbox{\XIU}  \pm \SXIU    $, \\
    \etau & $\mbox{\ETAU} \pm \SETAU   $, &
    \xdu  & $\mbox{\XDU}  \pm \SDELTAU $,
  \end{tabular}
\end{center}
where the value of \etau\ has been constrained using the published
OPAL measurements of the leptonic branching ratios and the \Pgt\ 
lifetime.  Limits on non-standard coupling constants and on the masses
of new gauge bosons are obtained.  The results are in agreement with
the V$-$A prediction of the Standard Model.
\bigskip\bigskip\bigskip\bigskip
\bigskip\bigskip
\begin{center}{\large
(Submitted to  The European Physical Journal C)
}\end{center}
\end{titlepage}
\begin{center}{\Large        The OPAL Collaboration
}\end{center}\bigskip
\begin{center}{
K.\thinspace Ackerstaff$^{  8}$,
G.\thinspace Alexander$^{ 23}$,
J.\thinspace Allison$^{ 16}$,
N.\thinspace Altekamp$^{  5}$,
K.J.\thinspace Anderson$^{  9}$,
S.\thinspace Anderson$^{ 12}$,
S.\thinspace Arcelli$^{  2}$,
S.\thinspace Asai$^{ 24}$,
S.F.\thinspace Ashby$^{  1}$,
D.\thinspace Axen$^{ 29}$
% ...
}\end{center}\bigskip
\bigskip
$^{  1}$School of Physics and Astronomy, University of Birmingham,
Birmingham B15 2TT, UK
\newline
$^{  2}$Dipartimento di Fisica dell' Universit{\`a} di Bologna and INFN,
I-40126 Bologna, Italy
\newline
$^{  3}$Physikalisches Institut, Universit{\"a}t Bonn,
D-53115 Bonn, Germany
\newline
$^{  4}$Department of Physics, University of California,
Riverside CA 92521, USA
\newline
$^{  5}$Cavendish Laboratory, Cambridge CB3 0HE, UK
\newline
$^{  6}$Ottawa-Carleton Institute for Physics,
Department of Physics, Carleton University,
Ottawa, Ontario K1S 5B6, Canada
\newline
$^{  7}$Centre for Research in Particle Physics,
Carleton University, Ottawa, Ontario K1S 5B6, Canada
\newline
$^{  8}$CERN, European Organisation for Particle Physics,
CH-1211 Geneva 23, Switzerland
\newline
$^{  9}$Enrico Fermi Institute and Department of Physics,
University of Chicago, Chicago IL 60637, USA
\newline
$^{ 10}$Fakult{\"a}t f{\"u}r Physik, Albert Ludwigs Universit{\"a}t,
D-79104 Freiburg, Germany
\newline
$^{ 11}$Physikalisches Institut, Universit{\"a}t
Heidelberg, D-69120 Heidelberg, Germany
\newline
$^{ 12}$Indiana University, Department of Physics,
Swain Hall West 117, Bloomington IN 47405, USA
\newline
$^{ 13}$Queen Mary and Westfield College, University of London,
London E1 4NS, UK
\newline
$^{ 14}$Technische Hochschule Aachen, III Physikalisches Institut,
Sommerfeldstrasse 26-28, D-52056 Aachen, Germany
\newline
$^{ 15}$University College London, London WC1E 6BT, UK
\newline
$^{ 16}$Department of Physics, Schuster Laboratory, The University,
Manchester M13 9PL, UK
\newline
$^{ 17}$Department of Physics, University of Maryland,
College Park, MD 20742, USA
\newline
$^{ 18}$Laboratoire de Physique Nucl{\'e}aire, Universit{\'e} de Montr{\'e}al,
Montr{\'e}al, Quebec H3C 3J7, Canada
\newline
$^{ 19}$University of Oregon, Department of Physics, Eugene
OR 97403, USA
\newline
$^{ 20}$Rutherford Appleton Laboratory, Chilton,
Didcot, Oxfordshire OX11 0QX, UK
\newline
$^{ 22}$Department of Physics, Technion-Israel Institute of
Technology, Haifa 32000, Israel
\newline
$^{ 23}$Department of Physics and Astronomy, Tel Aviv University,
Tel Aviv 69978, Israel
\newline
$^{ 24}$International Centre for Elementary Particle Physics and
Department of Physics, University of Tokyo, Tokyo 113, and
Kobe University, Kobe 657, Japan
\newline
$^{ 25}$Institute of Physical and Environmental Sciences,
Brunel University, Uxbridge, Middlesex UB8 3PH, UK
\newline
$^{ 26}$Particle Physics Department, Weizmann Institute of Science,
Rehovot 76100, Israel
\newline
$^{ 27}$Universit{\"a}t Hamburg/DESY, II Institut f{\"u}r Experimental
Physik, Notkestrasse 85, D-22607 Hamburg, Germany
\newline
$^{ 28}$University of Victoria, Department of Physics, P O Box 3055,
Victoria BC V8W 3P6, Canada
\newline
$^{ 29}$University of British Columbia, Department of Physics,
Vancouver BC V6T 1Z1, Canada
\newline
$^{ 30}$University of Alberta,  Department of Physics,
Edmonton AB T6G 2J1, Canada
\newline
$^{ 31}$Duke University, Dept of Physics,
Durham, NC 27708-0305, USA
\newline
$^{ 32}$Research Institute for Particle and Nuclear Physics,
H-1525 Budapest, P O  Box 49, Hungary
\newline
$^{ 33}$Institute of Nuclear Research,
H-4001 Debrecen, P O  Box 51, Hungary
\newline
$^{ 34}$Ludwigs-Maximilians-Universit{\"a}t M{\"u}nchen,
Sektion Physik, Am Coulombwall 1, D-85748 Garching, Germany
\newline
\bigskip\newline
$^{  a}$ and at TRIUMF, Vancouver, Canada V6T 2A3
\newline
$^{  b}$ and Royal Society University Research Fellow
\newline
$^{  c}$ and Institute of Nuclear Research, Debrecen, Hungary
\newline
$^{  d}$ and Department of Experimental Physics, Lajos Kossuth
University, Debrecen, Hungary
\newline
$^{  e}$ on leave of absence from the University of Freiburg
\newline

\clearpage
\section{Introduction}

A measurement of the Michel parameters in \Pgt\ decays is presented
which involes a novel method to fit the energy spectra and
energy-energy correlations of the charged decay leptons.
The \Pgt-pair data set used was produced in \Pep\Pem\ collisions at
center-of-mass energies close to $\sqrt{s}=m_\PZz$, corresponding to
an integrated luminosity of \Lumi.
The parameters are fitted to the lepton spectra of three event
classes, electron--hadron, muon--hadron and electron--muon, depending
on the decay modes of the two \Pgt\ decays in the event.
Previous measurements of the Michel parameters in \Pgt\ decays exist
that were performed at LEP~\cite{Aleph:michel,L3:michel} and at other
\Pep\Pem\ 
colliders~\cite{Argus:first,Argus:two-leptonic,Argus:eta,Argus:xi-delta,Argus:rho-tag,SLD:michel,Cleo:first,Cleo:michel,Crystal-Ball}.
Unlike previous measurements the presented analysis makes use of a
binned maximum likelihood fit of fully simulated events to the data.
It accounts for radiative corrections, detector effects, background
processes and selection efficiencies in a comparatively original way.

\subsection{Lorentz structure}

In the Standard Model the charged weak interaction is described by the
exchange of left-handed \PW\ bosons, i.e., by a pure vector coupling
to only left-handed fermions. Thus, in the low-energy four-fermion
ansatz, the Lorentz structure of the charged current is predicted to
be of the type~{``V$-$A \ot\ V$-$A''}. With this formalism the
dominant features of nuclear $\beta$ decay and of \Pgm\ decay are
correctly described~\cite{Fetscher:muon-decay}.  Deviations from this
behavior would indicate new physics and might be caused by changes in
the \PW-boson couplings or through interactions mediated by new gauge
bosons~\cite{Pich:new-interactions}.  The dominant contribution would
then be either the one with the largest coupling or the one being
mediated by the lightest boson. This means that if there exist
contributions to the leptonic decay structure other than V$-$A, for
example a right-handed vector coupling arising from a heavy
right-handed \PW\ boson, \PWR, or a scalar coupling to a charged Higgs
boson, these would emerge first in the decay of the massive \Pgt\ 
lepton.  Among all its decay modes, the decays \tautoe\ and \tautom\ 
are the only ones in which the electroweak couplings can be probed
without disturbance from the strong interaction.  This makes the
purely leptonic \Pgt\ decays an ideal system to study the Lorentz
structure of the charged weak current.

The leptonic decay amplitude can be generalized by adding
current-current terms for all possible bilinear covariants. The most
general, derivative-free, four-lepton interaction matrix element for
the \tautol\ decay that is local and Lorentz invariant can be written
as (see e.g.~\cite{Scheck:muon-physics}):
\begin{equation}
{\cal M} = 4\frac{G_0}{\sqrt{2}}
\sum_{\stackrel{\gamma=\mathrm{S,V,T}}{\epsilon,\omega=\mathrm{R,L}}}
g_{\epsilon\omega}^{\gamma} \,
\langle \, \Pal_\epsilon|\Gamma^\gamma|\Pgnl \rangle \,
\langle \Pagngt|\Gamma_\gamma|\Pgt_\omega \rangle.
\label{eqn:matrix-element}
\end{equation}
Here $\gamma$ denotes the type of the interaction (scalar, vector or
tensor) and $\Gamma^{\gamma}$ are $4\times 4$ matrices defined in
terms of the Dirac matrices:
\begin{equation}
\Gamma^{\rm S} = 1, \qquad \Gamma^{\rm V} = \gamma^\mu, \qquad
\Gamma^{\rm T} = \frac{1}{\sqrt{2}}\,\sigma^{\mu\nu}
\equiv \frac{i}{2\sqrt{2}} (\gamma^\mu\gamma^\nu - \gamma^\nu\gamma^\mu).
\end{equation}
The indices $\omega$ and $\epsilon$ denote the chiralities of the
\Pgt\ lepton and its charged decay lepton, \Pl, respectively. For
given ($\omega$, $\epsilon$) the chiralities of the neutrinos are
uniquely determined.  Tensor interactions exist only for opposite
chiralities of the charged leptons. This leads to 10 complex coupling
constants, $g_{\epsilon\omega}^{\gamma}$,
for which the Standard Model predicts \gvll=1 and all others being
zero. Choosing the arbitrary phase by defining \gvll\ to be real and
positive leaves 19 real numbers to be determined by experiment.  As
long as one is interested in the relative strengths of the couplings,
it is convenient to require the following normalization condition:
\begin{equation}
N \equiv
\renewcommand{\arraystretch}{1.5}
\begin{array}[t]{rl}
\frac{1}{4}\! & \left( {|\gsll|}^2 + {|\gslr|}^2
                     + {|\gsrl|}^2 + {|\gsrr|}^2 \right)\\
          +\! & \left( {|\gvll|}^2 + {|\gvlr|}^2
                     + {|\gvrl|}^2 + {|\gvrr|}^2 \right)\\
       +\ 3\! & \left( {|\gtlr|}^2 + {|\gtrl|}^2 \right) = 1 .
\end{array}
\label{eqn:normalization}
\end{equation}
\label{sec:ranges}
This restricts the allowed ranges of the coupling constants to
$|\gs|\leq 2$, $|\gv|\leq 1$ and $|\gt|\leq \frac{1}{\sqrt{3}}$. The
overall normalization can be incorporated into $G_0$ which then
accounts for deviations from the Fermi constant \gfermi.

\subsection{Michel parameters}
\label{sec:michel-parameters}

At the Born level, neglecting radiative corrections and terms
proportional to $(\frac{m_\Pl}{m_\Pgt})^2$, only four different
combinations of these coupling constants, denoted by $\rho$, $\xi$,
$\delta$ and $\eta$, determine the shape of the decay spectra. In the
\Pgt\ rest frame, the leptonic decay width can, for massless
neutrinos, be written as:
%
\begin{eqnarray}
\frac{\dd^2\Gamma_{\tautol}}{\dd\Omega\,\dd{x^*}} & = &
\frac{G_0^2 m_\Pgt^5}{192\pi^4} \, {x^*}^2
\left\{
        3(1-{x^*}) + \mathbf{\rhol}
        \left( \frac{8}{3}{x^*}-2 \right)
        + 6 \, \mathbf{\etal} \frac{m_\Pl}{m_\Pgt}
        \frac{(1-{x^*})}{x^*}
\right. \nn \\
& & \left.
        - \Ptau \, \mathbf{\xil} \cos{\theta^*}
        \left[
                (1-{x^*})+\mathbf{\deltal}
                \left( \frac{8}{3}{x^*}-2 \right)
        \right]
\right\}.
\end{eqnarray}
Here ${x^*}=\frac{{E_\Pl}^*}{{E_\Pl^\mathrm{max}}}$ is the scaled
energy of the charged decay lepton in the \Pgt\ rest frame with
${{E_\Pl^\mathrm{max}}}=\frac{m_\Pgt^2+m_\Pl^2}{2m_\Pgt}$ being its
maximal energy and $\cos{\theta^*}$ is the angle between the \Pgt-spin
direction and the momentum of the decay lepton. \Ptau\ is the average
\Pgt\ polarization. After integration over $\cos{\theta^*}$ and
boosting into the \PZz\ rest frame, the spectrum has the form:
\begin{equation}
\label{eqn:single-decay}
H(x) = f(x) + \Ptau\, g(x), \qquad\mathrm{with}
\end{equation}
\begin{equation}
f(x) = a(x)+\rho\,b(x)+\eta\,e(x), \qquad
g(x) = \xi\,c(x)+\xi\delta\,d(x),
\end{equation}
where $f(x)$ and $g(x)$ describe the isotropic and the
\Pgt-spin-dependent part, respectively, and $a(x) \dots e(x)$ are
known third-order polynomials in the scaled energy $x =
\frac{E_\Pl}{E_\Pgt}$.
Hence, allowing the most general couplings, the shape of the spectra
can be described by the four Michel parameters~\cite{Michel,Bouchiat}
for which the Standard Model predicts the values $\rho=\frac{3}{4}$,
$\xi=1$, $\delta=\frac{3}{4}$ and $\eta=0$ according to a V$-$A
structure of the charged weak current.
Their definitions read in detail:\footnote{Since by definition
  $\delta$ contains a factor $\frac{1}{\xi}$ it is convenient to use
  $(\xi\delta)$ instead of $\delta$.}
\setlength{\jot}{3\jot}
\begin{eqnarray}
  \rho & = &
    \frac{3}{4}  {|\gvll|}^2 + \frac{3}{4}  {|\gvrr|}^2
  + \frac{3}{16} {|\gsll|}^2 + \frac{3}{16} {|\gslr|}^2
  + \frac{3}{16} {|\gsrl|}^2 + \frac{3}{16} {|\gsrr|}^2
  \nn \\
  &   &
  + \frac{3}{4} {|\gtlr|}^2 + \frac{3}{4}   {|\gtrl|}^2
  - \frac{3}{4} \mathrm{Re}(\gslr \gtlrs)
  - \frac{3}{4} \mathrm{Re}(\gsrl \gtrls)
  \nn \,, \\
  \xi & = &
  {|\gvll|}^2 + 3{|\gvlr|}^2 - 3{|\gvrl|}^2 - {|\gvrr|}^2
  + 5{|\gtlr|}^2 - 5{|\gtrl|}^2
  \nn \\
  &   &
  + \frac{1}{4}{|\gsll|}^2 - \frac{1}{4}{|\gslr|}^2 
  + \frac{1}{4}{|\gsrl|}^2 - \frac{1}{4}{|\gsrr|}^2
  + 4\mathrm{Re}(\gslr\gtlrs) - 4\mathrm{Re}(\gsrl\gtrls) 
  \,, \\
  \xi\delta & = &
    \frac{3}{4} {|\gvll|}^2 - \frac{3}{4} {|\gvrr|}^2
  + \frac{3}{16}{|\gsll|}^2 - \frac{3}{16}{|\gslr|}^2
  + \frac{3}{16}{|\gsrl|}^2 - \frac{3}{16}{|\gsrr|}^2
  \nn \\
  &   &
  - \frac{3}{4} {|\gtlr|}^2 + \frac{3}{4} {|\gtrl|}^2
  + \frac{3}{4}\mathrm{Re}(\gslr\gtlrs)
  - \frac{3}{4}\mathrm{Re}(\gsrl\gtrls)
  \nn \,, \\
  \eta & = &
  \frac{1}{2} \mathrm{Re} \left[
    \gvll\gsrrs +  \gvrr\gslls 
    + \gvrl (\gslrs + 6 \gtlrs) + \gvlr (\gsrls + 6 \gtrls)
  \right]
  \nn .
\label{eqn:michel-parameters}
\end{eqnarray}
One of the parameters, $\eta$, can also lead to a change in the
leptonic decay width of the \Pgt\ lepton.

If contributions from other couplings exist, they are not necessarily
the same for \tautoe\ and \tautom.  Therefore, the Michel parameters
for decays into electron and muon are measured independently. Because
the \etal-term is suppressed by a factor $\frac{m_\Pl}{m_\Pgt}$, there
is almost no sensitivity to \etae\ from the \tautoe\ decay spectrum.
Thus in the following \etae\ is set to zero. This leaves the 7
parameters \rhoe, \xie, \xde, and \rhom, \xim, \xdm, \etam\ to be
determined.

To test the Standard Model prediction the parameters will also be
fitted under the assumption of \Pe-\Pgm\ universality. It is possible
to test explicit extensions, for example by focusing on vector
couplings, or by allowing only one additional scalar contribution. In
this way, mass limits for a right-handed \PW\ boson, \PWR, as well as
for a charged Higgs boson are determined.

\subsection{Lepton-lepton correlations}
\label{sec:correlation-function}

At LEP, \Pgt\ pairs are produced with almost perfect spin
correlation.\footnote{For V and A type couplings in the production,
  the \Pgtp\ and \Pgtm\ have opposite chiralities.}
This allows one to measure the spin-dependent part of the decay
spectra with high sensitivity by employing the correlations between
both \Pgt\ decays in the
event~\cite{Nelson:correlation-functions,Fetscher:leptonic-tau-decays}.
For parallel \Pgt\ spins the correlation function can be written as:
\begin{equation}
I(x_1,x_2) = f(x_1) f(x_2) + g(x_1) g(x_2)
  - \Ptau \left[\, f(x_1) g(x_2) + f(x_2) g(x_1) \,\right],
\end{equation}
where $f(x)$ and $g(x)$ are the above third-order polynomials and
\Ptau\ is the average polarization of the \Pgtm\ lepton.
In contrast to uncorrelated single decay spectra
(eq.~\ref{eqn:single-decay}) here the product of the two
spin-dependent parts, $g(x_1) g(x_2)$, without the suppression by the
\Pgt-polarization.  Thus, in case of two leptonic decays in the event,
the correlation function provides high sensitivity to the parameters
$\xi$ and $\xi\delta$ whereas the more frequent events with single
leptonic decays contribute with high statistics to the measurement of
$\rho$ and $\eta$.

\subsection{Tau polarization}

In addition to the charged current couplings which determine the decay
of the \Pgt\ lepton, the couplings to the neutral current which are
responsible for \Pgt\ production influence the shape of the spectra.
The difference in the left- and right-handed \PZz\ couplings causes
the \Pgt\ leptons to be produced polarized affecting the
spin-dependent part of the decay spectra.  In principle, the average
\Pgt-polarization, \Ptau, could be measured along with the Michel
parameters. However, this introduces additional correlations between
the fit parameters and limits the accuracy of the Michel parameter
measurements while it does not reveal anything new about the decay
structure.  Since extensions to the charged sector of the weak
interaction do not a priori change the neutral current, a different
approach is pursued in this analysis. Instead of testing both sectors
simultaneously at the expense of less sensitivity, the charged current
is investigated in the most general way, while the neutral current
(i.e., the \Pgt-polarization) is assumed to be described by the
Standard Model couplings with adequate accuracy.

Since all direct measurements of \Ptau\ from \eett\ data at
$\sqrt{s}=m_\PZz$ implicitly have assumed a V$-$A coupling in the
\Pgt\ decay, the use of these results as input to this analysis would
introduce a bias. Therefore, measurements of the neutral current which
are independent of the charged sector have been used to calculate
\Ptau\ using the \ZFITTER~\cite{Zfitter} package. The uncertainty
arising from this procedure has been studied with the systematic
errors. Its impact on the Michel parameter measurement is smaller than
the one that would have been introduced through correlations between
the parameters and the polarization if \Ptau\ were fitted.
  
As input to the \Ptau\ calculation the following values were
used~\cite{LEP:EW}: the preliminary LEP measurement of the \PZz\ mass
$m_\PZz = \MZ$, the CDF/D$\emptyset$ combined value for the top mass
$m_\Pt = \Mt$, $m_\PHz = 300\,\GeV$ with $60\,\GeV < m_\PHz <
1\,\TeV$, $\alphas(m_\PZz) = \Alphas$ and $\alpha(m_\PZz)^{-1} =
\Alpha$.
These values yield the predicted value of the \Pgt-polarization at the
\PZz\ peak as $\Ptau = -0.1391 {+0.0069\atop-0.0055}$, where the
dominant uncertainty is due to the unknown Higgs boson mass.  In this
analysis, data taken at $\sqrt{s} = m_\PZz$ are used together with
data below and above $m_\PZz$.  Averaging over the energy dependence
of \Ptau\ according to the data set used yields the same quoted value.
Over the full range this leads to an uncertainty in \Ptau\ of
approximately $5\,\%$ which is still smaller than the error of the
current measurements (see e.g.~\cite{tau-polarization-96}).

\section{Event selection}

\subsection{OPAL detector}

The OPAL detector is described in detail
elsewhere~\cite{Opal-detector,Opal-silicon}. Here, only a brief
summary of the main components shall be given.
The innermost subdetector is a micro-vertex detector with two layers
of double-sided silicon-strips.
It is enclosed by a system of three different drift chambers, a
precision vertex chamber with axial and stereo wire readout, a large
cylindrical drift volume (jet chamber) with 24 azimuthal sectors
of 159 signal wires each, and a surrounding set of $z$-chambers with
wires perpendicular to the beam direction.
The central tracking system is contained inside a magnetic coil which
provides a solenoidal field of 0.435 Tesla. This leads to a resolution
of the transverse momentum of $\sigma_{p_t}/p_t\approx
1.5\cdot10^{-3}\cdot p_t\,(\GeV)$.  In addition, the specific energy
loss, $\dd E/\dd x$, of charged particles is measured in the jet
chamber.
The electromagnetic calorimeter (ECAL) is built of 11,704 lead glass
blocks of approximately $25\,X_0$ radiation lengths, providing an
energy resolution of typically $\sigma_E/E\approx
12\,\%/\sqrt{E\,(\GeV)}$.
In front of the ECAL, a thin gas detector (the presampler) measures
electromagnetic showers beginning in the material of the inner
detector components.
The ECAL is surrounded by the iron return yoke of the magnet which is
instrumented with limited streamer tubes to serve as a hadron
calorimeter (HCAL).
The whole detector is enclosed by 4 layers of muon chambers, giving
the position and the direction of the penetrating particles.

As observables for the fit, different variables in the two leptonic
decay channels are chosen. In case of the \tautom\ decays, the track
momentum of the muon measured in the jet chamber, $p_\mathrm{track}$,
is taken, whereas for the \tautoe\ decays the energy deposited by the
electron in the electromagnetic calorimeter, $E_\mathrm{cluster}$, is
used.  The cluster energy includes a large fraction of final state
photon radiation from the electron and is thus less dependent on the
modeling of the photon radiation in the generator and detector
simulation.  Both variables are scaled to the beam energy which is
used as the estimator for the maximal energy of the decay lepton.

\subsection{Tau pair selection}

\label{sec:tau-pair-selection}

From the data collected with the OPAL detector during the years
1990--1995, \Pgt-pair events are selected in several steps. The
selection follows the strategy described in earlier OPAL publications
\cite{tau-polarization-96,tau-polarization-95} and given in detail in
\cite{tau-polarization-91}.  First, lepton pairs are preselected by
requiring exactly two charged jet-cones of $35^\circ$ half-opening
angle with low track and cluster multiplicity.  Two-photon events are
rejected by requiring either a large visible energy or an unbalanced
transverse momentum sum, and a small acollinearity.

From this sample Bhabha events are removed based on a large sum of
cluster energies or a large sum of track momenta in conjunction with
large cluster energies. Afterwards, \Pgm-pair events are eliminated if
consistent with high track momenta, small energy deposit in the
calorimeter or signals in the muon chambers. The remaining \Ntaupairs\ 
events are almost entirely \Pgt\ pairs (see
section~\ref{sec:background} for the remaining background in the used
event classes).  The geometrical acceptance of this selection covers
the region $|\cos\theta| < 0.95$.

\subsection{Tau decay mode identification}

\label{sec:likelihood-identification}

A likelihood selection is used to identify the \Pgt-decay modes in the
two cones. It distinguishes between the 1-prong decays \tautoe,
\tautom\ and \tautoh\ where \Ph\ is either \Pgp/\PK, \Pgr\ or
$\Pai\ra\Pgp 2\Pgpz$.  It makes use of a set of variables which allow
the discrimination of different channels.\footnote{Kaons are not
  separated from pions and are counted among the corresponding
  \Pgp-channels in the following.}  These variables include:

\begin{list}{}{\labelwidth 10em \leftmargin 8em }
\item[$E_\mathrm{cluster}/p_\mathrm{track}$ --] the ratio of the
  energy deposited in the electromagnetic calorimeter (ECAL) with
  respect to the track momentum measured in the central detector,
  
\item[$\dd E/\dd x$ --] the specific energy loss in the jet chamber,
  
\item[$N_\mathrm{blocks}^{90\,\%}$ --] the number of ECAL blocks that
  contain 90\,\% of the measured energy in the cluster associated to
  the track,
  
\item[$E_\mathrm{neutral}/E_\mathrm{cone}$ --] the fraction of ECAL
  energy in the cone that is not associated to the track,
  
\item[$(\Delta\phi)_\mathrm{max}$ --] the maximum angle between the
  track and a presampler cluster assigned to the cone,
  
\item[$W_\mathrm{pres}$ --] the width of the largest presampler
  cluster,
  
\item[$N_\mathrm{hits}/N_\mathrm{layers}$ --] the average number of
  hits per active layer in the hadronic calorimeter (HCAL),
  
\item[$N_\mathrm{7 hits}$ --] the number of hits in the last 3 HCAL
  layers and in the 4 layers of the muon chambers,

\item[$W_\mathrm{muon}$ --] the matching probability between the
  extrapolated track and a muon chamber track segment.
\end{list}
The measured variables are then compared to a set of reference
distributions that have been produced by a Monte Carlo simulation of
the considered decay modes (see section \ref{sec:monte-carlo}).

Based on the variable $x_i$, the expected fraction of decay modes of
the type $j$ is given as:
\begin{equation}
  \ell_i^j(x_i) = \frac{f_i^j(x_i)}{\sum_{j=1}^{N_\mathrm{modes}}
    f_i^j(x_i)} \,,
\end{equation}
where $f_i^j$ are the normalized probability densities taken from the
reference distribution for the respective variable $i$ and
$N_\mathrm{modes} = 5$ is the number of considered decay modes. The
information from all the variables is then combined into the product
likelihood for the hypothesis $j$:
\begin{equation}
  {\cal L}^j(x_1,\dots,x_{N_\mathrm{variables}}) =
  \prod_{i=1}^{N_\mathrm{variables}} \ell_i^j(x_i) \,.
\end{equation}
Normalized to all considered alternatives the expression yields the
relative likelihood of the decay being of type $j$:
\begin{equation}
  P^j = \frac{{\cal L}^j}{\sum_{j=1}^{N_\mathrm{modes}} {\cal L}^j} \,,
\end{equation}
which in the case of uncorrelated variables can be interpreted as a
probability.
To obtain high purity samples, the cones involving the lepton are
required to be identified with a relative likelihood of $P^j >
90\,\%$.
The simulated reference distributions are checked against the data
using a tagging technique. For that purpose, the likelihood variables
of each detector component are compared between data and Monte Carlo
based on a clear decay mode identification from the other components.
A comparison of some of the variables can be found in
\cite{CP-violation}.

\subsection{Selection efficiency}

Throughout the analysis, the simulated events and the data are treated
in an identical manner. Thus, the efficiencies for the event selection
and the decay mode identification are accounted for in the fit as
modeled in the Monte Carlo.
The validity of the simulation has been tested by comparing Monte
Carlo efficiencies as a function of energy / momentum with data
control samples (see section \ref{sec:systematic-errors}).
Table~\ref{tab:taupid-effi} lists the efficiencies for the decay mode
identification within the fiducial volume of $|\cos\theta| < 0.95$.
The inefficiency of the muon channel originates mainly from the low
$x$ region where the separation from pions becomes poorer.

\begin{table}[hbt]
\centering
\begin{tabular}{l|ccccc}
               & \multicolumn{5}{c}{Identified mode} \\
Generated mode & \tautoe & \tautom & \tautop & \tautor & \tautoa \\
\hline
\tautoe & $96.5$ & $  - $ & $ 1.1$ & $ 1.7$ & $ 0.7$ \\
\tautom & $ 0.3$ & $92.2$ & $ 7.0$ & $ 0.5$ & $  - $  \\
\tautop & $ 1.4$ & $ 3.3$ & $82.1$ & $ 9.5$ & $ 3.6$ \\
\tautor & $ 0.3$ & $ 0.2$ & $10.3$ & $67.8$ & $21.3$ \\
\tautoa & $ 0.1$ & $  - $ & $ 1.3$ & $25.8$ & $72.7$ \\
\end{tabular}
\caption{\em 
Efficiencies in \% for the likelihood identification of 1-prong \Pgt\ decay
modes within the fiducial region.  The diagonal values are the
efficiencies for the true identified modes whereas the off-diagonal
values represent the misidentification probabilities.
}
\label{tab:taupid-effi}
\end{table}

Based on the decays of both \Pgt\ leptons, the events are divided into
mutually exclusive samples: lepton-lepton correlations (\Pe--\Pgm) and
single-lepton decays (\Pe--hadrons, \Pgm--hadrons).  As mentioned in
section~\ref{sec:correlation-function}, the correlation spectra have a
high sensitivity to the spin-dependent parameters (\xiu, \xdu) while
the single decays provide high statistics for the isotropic parameters
(\rhou, \etau).

Decay correlations in \Pe--\Pe\ and \Pgm--\Pgm\ events are subject to
large background contamination from \eeee\ and \eemm\ processes as
well as from two-photon events \ggee\ and \ggmm. These background
sources distort the correlation spectra in the two most sensitive
regions where both leptons have high energies or both have low
energies, respectively.  Due to the uncertainties in the estimation of
these backgrounds, the \Pe--\Pe\ and \Pgm--\Pgm\ event classes would
contribute with a dominant systematic error to the fit result. For
this reason they are not included in this analysis.

In table~\ref{tab:spec-effi} the overall efficiencies for the three
event classes are given.  The additional inefficiencies with respect
to the decay mode identification arise from geometrical cuts that are
performed on the individual event classes to exclude insensitive or
inadequately simulated regions of the detector. These regions include
the proximity of the anode wire planes of the jet chamber,
$\phi_\mathrm{sector} < 0.4^\circ$, the extreme forward region of the
electromagnetic calorimeter, $|\cos\theta| > 0.9$, and a few small
regions not covered by the muon chambers.

\begin{table}[hbt]
\centering
\begin{tabular}{c|ccc}
& \multicolumn{3}{c}{Event Class} \\
& \tauitoe{1} & \tauitom{1} & \tauitoe{1} \\
& \tauitoh{2} & \tauitoh{2} & \tauitom{2} \\
\hline
Sample size & \Neh   & \Nmh   & \Nem    \\
Efficiency  & \EFFeh & \EFFmh & \EFFem \\
\end{tabular}
\caption{\em Selection efficiencies for the three considered event classes.}
\label{tab:spec-effi}
\end{table}

\begin{figure}[p]
  \centering
  \resizebox{0.6\textwidth}{!}{\includegraphics{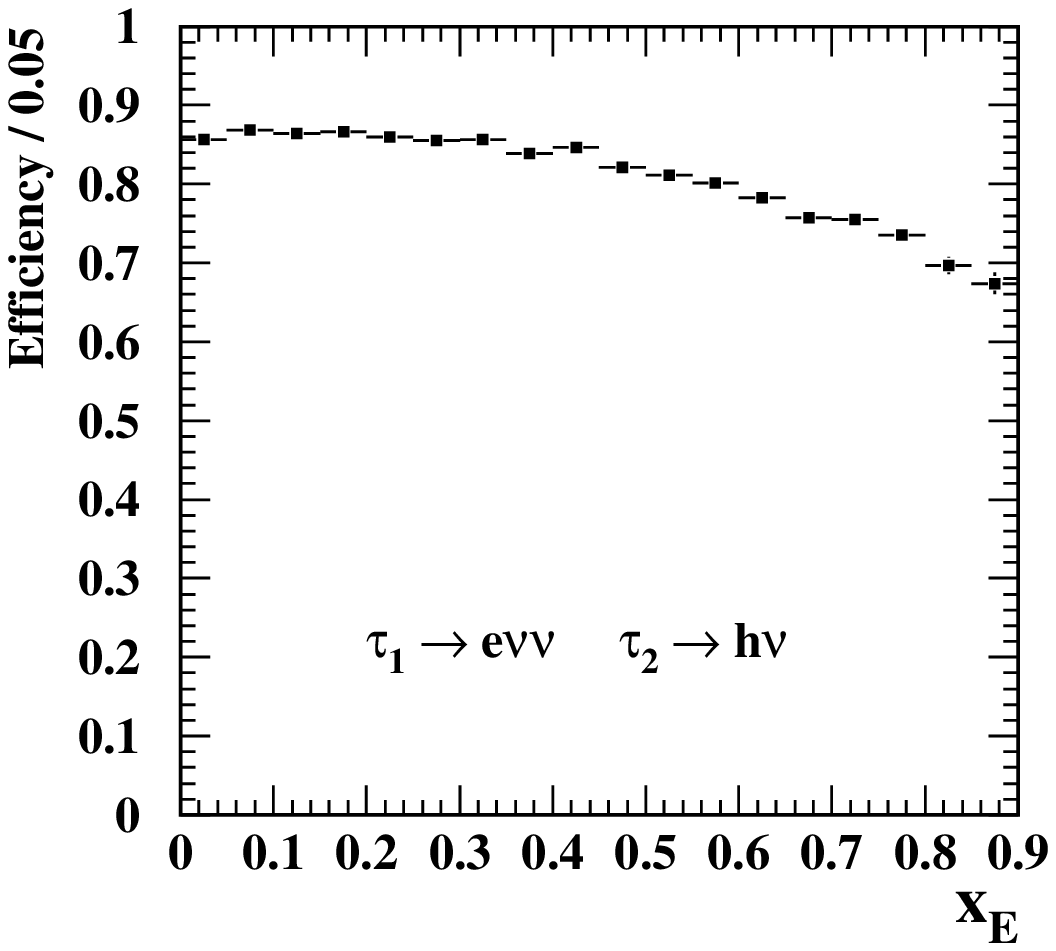}}
  \resizebox{0.6\textwidth}{!}{\includegraphics{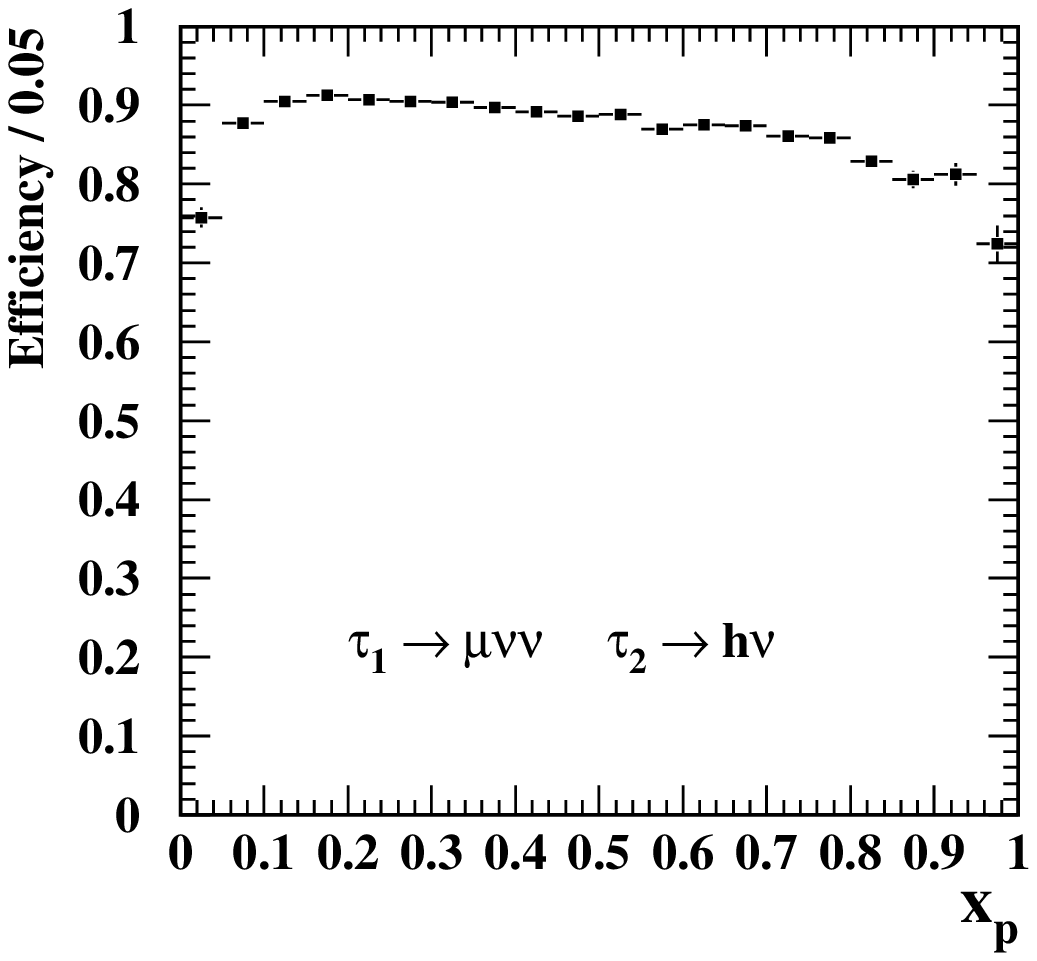}}
\caption{\em 
  Selection efficiencies for the \Pe-\Ph\ (\Pgm-\Ph) event samples as
  a function of the scaled lepton energy (momentum), $x_E$ ($x_p$), as
  determined from the Monte Carlo after the preselection including all
  fiducial cuts.}
\label{fig:single_effi}
\end{figure}

Figures~\ref{fig:single_effi}(a) and \ref{fig:single_effi}(b) show the
selection efficiencies for the single leptonic decay samples \Pe-\Ph\ 
and \Pgm-\Ph\ as a function of the scaled energy $x_E$ and scaled
momentum $x_p$, respectively.

\subsection{Background}

\label{sec:background}

Various sources of background in the \Pgt-pair sample and in the
\tautoe\ and \tautom\ samples have been investigated.  The main
contribution comes from misidentification of the hadronic \Pgt\ 
decays.  Due to the distinctive signatures of electrons and muons in
the detector, cross-talk between the two channels is negligible. The
main backgrounds for the \tautoe\ mode originate from \tautop\ and
\tautor\ decays.  For the \tautom\ mode the dominant background source
are \tautop\ decays. Table~\ref{tab:background} lists the different
backgrounds in the three event samples. For simplicity, the three
identified hadron channels have been summed up.  For the single decay
spectra, double lepton events in which the recoil lepton has been
misidentified as a hadron are also quoted as background.
\begin{table}[hbt]
\centering
\begin{tabular}{lc|ccc}
\multicolumn{2}{c|}{Background source} & \multicolumn{3}{c}{Event class} \\
               &             & \Pe --\Ph  & \Pgm --\Ph & \Pe --\Pgm \\
\hline
leptonic \Pgt\ & \Pe --\Pe   & \BKGehee   & $-$        & \BKGemee   \\
               & \Pe --\Pgm  & \BKGehem   & $-$        &            \\
               & \Pgm--\Pe   & $-$        & \BKGmhme   &            \\
               & \Pgm--\Pgm  & $-$        & \BKGmhmm   & \BKGemmm   \\
\hline       
hadronic \Pgt\ & \Ph --\Ph   &  \BKGehhh   & \BKGmhhh   & $-$       \\
               & \Pe --\Ph   &            & $-$        & \BKGemeh   \\
               & \Ph --\Pgm  & $-$        &            & \BKGemhm   \\
\hline      
non-\Pgt\      & \EEEE\      & \BKGehEE   & $-$        & \BKGemEE   \\
               & \EEMM\      & $-$        & \BKGmhMM   & \BKGemMM   \\
               & \ggEE\      & \BKGehEEEE & $-$        & \BKGemEEEE \\
               & \ggMM\      & $-$        & \BKGmhEEMM & \BKGemEEMM \\
\hline       
& \\
Total          &             & \BKGeh     & \BKGmh      & \BKGem    \\
\end{tabular}
\caption{\em Identified events and background from \Pgt\ decays, Bhabha,
  \Pgm-pair and two-photon events. For simplicity the three hadronic channels
   have been summed up.}
\label{tab:background}
\end{table}

Furthermore, Bhabha and \Pgm-pair events may pass the selection and
contaminate the samples. This can occur, for example, for an \eeee\ 
event where one electron is misidentified as \tautor\ decay, or to an
\eemm\ event where one muon fakes a \tautop\ decay.

As mentioned earlier, the presented method allows to account for all
backgrounds that are simulated in the Monte Carlo. Since either the
background shapes do not depend on the Michel parameters (hadronic and
non-\Pgt\ background) or their dependence has, due to the small
fraction, no significant impact on the shape of the signal
distribution, the residual events can be independently added to the
fit spectrum.

\section{Fitting method}

To extract the Michel parameters from the observed spectra, a binned
maximum likelihood fit of a set of Monte Carlo spectra has been
applied to the data.
Compared to fits that involve analytical functions to describe the
data distributions, this method has several advantages.  It includes
radiative corrections at the generator level where their description
is more easily accessible than it is through the convolution of photon
radiation probability functions. It provides a full simulation of the
detector response and thus accounts for the energy and momentum
resolution as well as for the selection efficiency in an elegant way.
It also accounts for contamination from misidentified \Pgt\ decays and
from other background sources, like Bhabha events, \Pgm-pair events
and two-photon events.  This makes it unnecessary to unfold resolution
and efficiency effects from the data.  A particular benefit is that
selection criteria which restrict the phase space do not present a
problem for the fit because all requirements can be placed identically
on data and Monte Carlo.  In addition, observables for each decay mode
can be chosen to accommodate the particular capabilities of the
detector rather than to facilitate the theoretical description.

\subsection{Linear combination}

Since the single decay spectra depend linearly on the Michel
parameters, it is possible to decompose any observed spectrum into
different basis spectra with each representing a specific set of
parameters.  To describe the whole parameter space including the
constant term, five spectra have to be mixed with coefficients that
add up to unity.  For a particular choice of the basis parameter sets,
the corresponding coefficients can be calculated by solving the
respective equation system. This allows one to determine the values of
the $4$ Michel parameters by fitting the coefficients with respect to
a given basis (i.e.\ the relative contributions of the basis spectra)
to the data.

In the case of the energy-energy correlation spectra between two
leptonic decays in each event (double decay spectra) this method is
still applicable. These spectra can be represented by a composition
that is bilinear in the $2\times4$ Michel parameters of the \Pgt\ 
decays into electron and muon (or of second order in the $4$
parameters if the leptons in both hemispheres are identical).  The
correlation basis is then the tensor product of two single decay
bases.  Now, 25 coefficients appear in the decomposition, but are not
all independent.  From these, the $2\times4$ Michel parameters for
\Pe\ and \Pgm\ decays, respectively, can again be calculated by matrix
inversion (see section~\ref{sec:parameter-basis}). In this formalism,
the \Pe-\Pe\ and \Pgm-\Pgm\ correlations with only $4$ free Michel
parameters are just a special case of the \Pe-\Pgm\ 
correlations.\footnote{They are formally analog to
  \Pe-\Pgm-correlations under the assumption of \Pe-\Pgm\ 
  universality.}  They will not be discussed explicitly since they are
not used for the analysis.

With this method, which is based on Monte Carlo event generation, it
is possible to describe any value of the Michel parameters with a
finite sample of events by varying the appropriate contribution to the
spectrum.

\subsection{Parameter basis}  
\label{sec:parameter-basis}

To describe the entire Michel parameter space, a basis has to be
chosen which accounts for any possible combination.  It is obvious
that the parameter sets corresponding to pure S,P,V,A,T couplings do
not fulfil this requirement because they do not involve any
interference terms between different couplings. Also, the canonical
basis with $(\rhou,\xiu,\xdu,\etau) = (1,0,0,0),(0,1,0,0),\dots$ etc.\ 
is not physically meaningful, because it is not possible to have
\xdu=1 and all other Michel parameters zero. However, it is desirable
to have a ``nearly orthogonal'' set in order to avoid large
correlations between the coefficients. Examples of different couplings
and the corresponding Michel parameters can be found in
table~\ref{tab:couplings}.

Taking this into consideration, the following basis has been chosen:
\begin{equation}
      \left( \begin{array}{c} \rhou \\ \xiu \\ \xdu \\ \etau \end{array} \right)
= c_1 \left( \begin{array}{c}  3/4  \\   1  \\  3/4 \\   0   \end{array} \right)
+ c_2 \left( \begin{array}{c}   1   \\   0  \\   0  \\   0   \end{array} \right)
+ c_3 \left( \begin{array}{c}   0   \\   1  \\   0  \\   0   \end{array} \right)
+ c_4 \left( \begin{array}{c}  3/4  \\   0  \\   0  \\  1/2  \end{array} \right)
+ c_5 \left( \begin{array}{c}   0   \\   0  \\   0  \\   0   \end{array} \right)
,
\end{equation}
with $\sum_i c_i = 1$. Here, each vector on the right-hand side stands
for a spectrum generated with the quoted set of Michel parameters.

In a more general notation\footnote{To simplify the notation,
  $\delta$ is written instead of \xdu\ in this section.}, one can
write for the case of the single lepton spectra:
\begin{equation}
\mathbf{q}_{\Pl} = M_{\Pl} \cdot \mathbf{c}_{\Pl},
\end{equation}
where $\mathbf{q}_{\Pl}$ is the vector $(1,\rhol,\xil,\deltal,\etal)$,
$\mathbf{c}_{\Pl}$ is the vector of coefficients $(c_1,c_2,c_3,c_4,c_5)$,
and $M_{\Pl}$ is a $5\times5$ matrix given by
\begin{equation}
M_{\Pl} =
\left(
\begin{array}{ccccc}
        1 &         1 &         1 &         1 &         1 \\
  \rhol_1 &   \rhol_2 &   \rhol_3 &   \rhol_4 &   \rhol_5 \\
   \xil_1 &    \xil_2 &    \xil_3 &    \xil_4 &    \xil_5 \\
\deltal_1 & \deltal_2 & \deltal_3 & \deltal_4 & \deltal_5 \\
  \etal_1 &   \etal_2 &   \etal_3 &   \etal_4 &   \etal_5
\end{array}
\right).
\end{equation}
Here $\rhol_1$ denotes the value of \rhol\ in the first basis
spectrum, $\rhol_2$ its value in the second spectrum and so forth.
Hence, for a given set of Michel parameters the coefficients with
respect to the basis $M_{\Pl}$ can be calculated from
\begin{equation}
\mathbf{c}_{\Pl} = {\left( M_{\Pl} \right)}^{-1} \cdot \mathbf{q}_{\Pl}.
\end{equation}
Analogously, for the case of the double \Pe-\Pgm\ spectra one can
write:
\begin{equation}
\mathbf{q}_{\Pe\Pgm} = M_{\Pe\Pgm} \cdot \mathbf{c}_{\Pe\Pgm},
\end{equation}
where the vector $\mathbf{q}_{\Pe\Pgm}$ is the outer product of the
vectors $\mathbf{q}_{\Pe}$ and $\mathbf{q}_{\Pgm}$ (
${q_{\Pe\Pgm}}_{ij} = {q_{\Pe}}_i \cdot {q_{\Pgm}}_j$ ),
$M_{\Pe\Pgm}$ is the corresponding outer product of the matrices
$M_{\Pe}$ and $M_{\Pgm}$, and $\mathbf{c}_{\Pe\Pgm} =
(c_1,\dots,c_{25})$ is the coefficient vector.
Again, for a given set of 8 Michel parameters the corresponding 25
coefficients in the actual basis are obtained by multiplying the
vector $\mathbf{q}_{\Pe\Pgm} =
(1,\rhoe,\xie,\deltae,\etae,\rhom,\rhoe\rhom,
\xie\rhom,\dots,\etae\etam) $ with the inverse of the $25\times25$
matrix $M_{\Pe\Pgm}$.


\subsection{Monte Carlo simulated samples} 
\label{sec:monte-carlo}

The \Pgt-pair Monte Carlo sample was generated using the
\KORALZ-3.8~\cite{Koralz-38} generator and a modified version of the
\TAUOLA-1.5~\cite{Tauola-15} decay library which was extended to
include the full generalized matrix
element~\cite{Schmitdler:matrix-element}. This allows the variation of
the Michel parameters over their whole range.

The \TAUOLA-1.5 version was preferred over the newer version
\TAUOLA-2.4~\cite{Tauola-15-update,Tauola-24} since it has a more
general approach to real photonic corrections.  The \TAUOLA-1.5 library
uses a factorization ansatz in order to produce real photons, where
each charged particle radiates independently of its decay matrix
element by internally applying the \PHOTOS~\cite{Photos} package. On
the other hand, the newer version \TAUOLA-2.4 contains full ${\cal
  O}(\alpha)$ corrections to take the interference between photons
radiated from the \Pgt\ and photons radiated from the decay lepton
into account. Although this might be more precise in the low energy
regime, it assumes a V$-$A type of interaction and is thus
model dependent. The former version, \TAUOLA-1.5, is capable of
producing photons for all possible decay structures at a reasonable
level of accuracy. It has therefore been chosen for this analysis.

Since finite Monte Carlo samples are used to describe possibly small
variations in the shape of the spectra, it is vital to keep
statistical fluctuations under control. If the distinct spectra were
generated independently their differences would be smeared by Gaussian
errors.  In particular, it could happen that in a certain bin, the
theoretical prediction for one spectrum is higher than for another but
the generated number of events is lower. To avoid such fluctuations,
each event is used for as many spectra as possible. This means that
most of the Monte Carlo spectra share a large fraction of common
events.  An acceptance/rejection method is used where an event is
flagged as accepted for each spectrum for which the generated random
weight is below the prediction (and not just the standard V$-$A) and
it is rejected only if it belongs to none of the considered spectra.
This procedure guarantees that the difference of any two spectra has
the right sign in all bins.  As a side effect, it makes the generation
of the Monte Carlo samples much more efficient.

Non-\Pgt\ background sources such as Bhabha, \Pgm-pair and two-photon
events have been generated using the \BABAMC~\cite{Babamc-1,Babamc-2},
\KORALZ~\cite{Koralz-40} and
\VERMASEREN~\cite{Vermaseren-1,Vermaseren-2} generators, respectively.
The response of the detector to generated particles is modeled using a
simulation program~\cite{GOPAL} based on the \GEANT~\cite{GEANT}
package. The simulated detector response has been checked with various
control samples (see section~\ref{sec:systematic-errors}).

\subsection{Fitting procedure}

The appropriate Monte Carlo distributions, each representing a basis
parameter set, are prepared with the identical binning as used for the
observed lepton spectra. The coefficients of the spectra,
$\mathbf{c}_\Pl$, are determined using a binned maximum likelihood fit
to the data.  To avoid any dependence on the description of the
overall efficiency, no constraint is made on the overall
normalization.  Since the coefficients are not independent, they are
not varied themselves, but are calculated from the corresponding
Michel parameters, as has been explained above.  Thus, the Michel
parameters are varied and fitted directly.  The actual minimization
and the determination of the covariance matrix is performed using the
\MINUIT~\cite{Minuit} package.  In doing this, a likelihood is
computed for every mixture, assuming Poisson errors in each bin.
Although the generated Monte Carlo sample is roughly four times larger
than the data sample, some spectra still have bins with only few
entries.  This occurs in particular in the correlation distributions
in regions of high $(x_1,x_2)$ as the various couplings may differ
drastically for extreme momentum configurations.  It is known that, in
the case of small bin entries, ignoring the Monte Carlo errors biases
the mean value of the fit and underestimates its spread.  Therefore,
fluctuations of both data and Monte Carlo have been taken into
account.
To accomplish this, an adjusted likelihood is
calculated~\cite{Barlow:MC-statistics} by finding in each bin $i$ the
most probable expectation with which data and Monte Carlo are
consistent:
\begin{equation}
\ln {\cal L} =
    \sum_{i=1}^{N_\mathrm{bins}}
    \sum_{j=1}^{N_\mathrm{spectra}}
        \left( a_{ji} \ln A_{ji} - A_{ji} \right)
  + \sum_{i=1}^{N_\mathrm{bins}} \left( d_i \ln f_i - f_i \right).
\label{eqn:adjusted_likelihood}
\end{equation}
Here $N_\mathrm{spectra}$ is the number of Monte Carlo basis spectra
(5 or 25), $d_i$ is the observed number of data events, $a_{ji}$ is
the generated number of Monte Carlo events in spectrum $j$ and
$A_{ji}$ is the best estimator for the Monte Carlo in the light of the
data.
The Monte Carlo expectation $ f_i = \sum_j p_j A_{ji} $ is the
composition of the best estimators using the mixing coefficients
$p_j$.
The first term of the adjusted likelihood
(\ref{eqn:adjusted_likelihood}) accounts for the agreement between the
actual Monte Carlo distribution ($a_{ji}$) and the ideal distribution
($A_{ji}$) for all spectra $j$. The second term accounts for the
agreement between the ideal composition ($f_i$) and the data ($d_i$).
The likelihood is maximized with respect to both the coefficients
$p_j$ and the estimators $A_{ji}$.\footnote{For each set of
  coefficients $p_j$ the numbers $A_{ji}$ can be calculated by solving
  an equation system.}
In the case of the double lepton spectrum, the index $i$ is replaced
by two indices. For clarity, all the following expressions correspond
to the single lepton spectra. The generalization to the double lepton
case is straightforward.

The above equation represents the correct treatment of the problem,
provided that the generated numbers, $a_{ji}$, are statistically
independent. As described before, this is not the case for the
prepared Monte Carlo spectra because they have (most) events in
common.  It is, however, possible to rewrite the expectation in each
individual bin by means of independent numbers.  To this end, the
spectra are ordered by increasing numbers of events in the considered
bin, and the coefficients are recalculated in terms of the differences
between the bins as follows.  If one writes for a particular
composition (setting $b_{0i} := 0$):
\begin{equation}
  f'_i = \sum_{j=1}^{N_\mathrm{spectra}} c_j\, b_{ji}
  = \sum_{j=1}^{N_\mathrm{spectra}} p_j\, ( b_{ji} - b_{j-1,i} ),
\end{equation}
where $b_{ji}$ is the generated number of events in bin $i$ of basis
spectrum $j$, then the differences $a_{ji} = b_{ji} - b_{j-1,i}$ are
all independent.  It follows that:
\begin{equation}
  p_j = \sum_{k=j}^{N_\mathrm{spectra}} c_k,
\end{equation}
where the coefficients $c_k$ are expressed in terms of the Michel
parameters as described above (section~\ref{sec:parameter-basis}):
\begin{equation}
  c_k = (M^{-1})_{kl} \cdot q_l.
\end{equation}
After this transformation the above expression for the likelihood is
applied.

\subsection{Constraint on the eta parameter}

It has been mentioned in section~\ref{sec:michel-parameters} that the
Michel parameter $\eta$ corresponds to a change in the partial decay
width when it deviates from zero. However, to eliminate any dependence
on the overall efficiency, Monte Carlo predictions are always
normalized to the data.  Due to the large correlation of $\eta$ with
the $\rho$ parameter, configurations are possible where the spectral
shape is changed only slightly while the value of $\eta$ is
inconsistent with the observed branching ratio.\footnote{ It has been
  shown that the leptonic branching ratios provide a sensitive
  observable on the Michel parameter $\eta$~\cite{Achim:eta}.}

Such a situation is avoided by constraining the branching ratios to
their measured values throughout the fit.
%
%
In the general ansatz for the Lorentz structure, the leptonic width is
in lowest order changed to:
\begin{equation}
  \Gamma_\ell(\etal) = \Gamma_\ell^\mathrm{(SM)}
  \left( 1 + 4 \etal\, \frac{m_\Pl}{m_\Pgt} \right),
\end{equation}
where $\Gamma_\ell^\mathrm{(SM)}$ is the Standard Model width (see
e.g.~\cite{Marciano-Sirlin}).
From the measurement of the \Pgt\ lifetime, $\tau_\Pgt$, the expected
branching ratio, $B_\ell \equiv B(\tautol)$, depending on the value of
\etal\ can be calculated as:
\begin{equation}
  B_\ell(\etal) = \Gamma_\ell(\etal)\,\tau_\Pgt.
\end{equation}
This relation can be used to calculate the most probable value of
$\eta$,
\begin{equation}
  \widehat{\etal} = \frac{1}{4} \frac{m_\Pgt}{m_\Pl}
  \left( \frac{B_\ell}
    {\Gamma_\ell^\mathrm{(SM)}\,{\tau_\Pgt}} - 1 \right).
\end{equation}
Using the published OPAL results for the \tautom\ branching ratio,
$B_\Pgm = \BRAmu$ \cite{tau-lifetime}, and the \Pgt\ lifetime,
$\tau_\Pgt = \TAUtau$ \cite{leptonic-branching-ratios}, one
determines:
\begin{equation}
  \widehat{\etam} = 0.032 \pm 0.073,
\end{equation}
which is consistent with zero.
The constraint to $\eta$ is applied by adding the following term to
the log likelihood:
\begin{equation}
  \ln {\cal L}^\mathrm{constraint}_\Pl = -\frac{1}{2}\,
  \frac{\left( \Gamma_\ell(\etal)\, \tau_\Pgt - B_\ell \right)^2}
  {\left( \Gamma_\ell(\etal)\, \Delta\tau_\Pgt \right)^2
    + \left( \Delta B_\ell \right)^2}.
\end{equation}
The use of this constraint makes the fit result for \etam\ dominated
by the branching ratio and lifetime measurements while the other
parameters are still sensitive to the allowed variation in the
\etam-dependent part of the shape.

\subsection{Checks of the fitting method}

To verify the reliability of the fit method, various checks have been
performed using the Monte Carlo event samples.  First, it has been
tested that the fit of the linear composition can reproduce the
genuine parameters of a specifically generated sample. It has also
been proven that any genuine spectrum and its corresponding mixture of
basis spectra are consistent within the statistical errors.

Second, the statistical errors of the fit parameters have been
checked. For that purpose, the Monte Carlo sample has been divided
into several pairs of subsamples with each pair representing a fit
sample and a fake data sample.  Then the fit has been performed for
each subset separately, and the distribution of the fit result has
been checked.  It has been verified that, when applying the adjusted
likelihood, the individual errors are consistent with the spread of
the mean values.

Third, it has been checked whether the particular choice of the
parameter basis affects the fit results. Other sets of basis spectra
than the quoted one have been used to describe a given Monte Carlo
distribution.  While some bases turned out to be less sensitive to one
or more of the parameters,
all alternatives have been found to be consistent within their fit
errors.

\section{Results}

\newcommand{\mspace}{\mbox{\rule[-1.8ex]{0ex}{5ex}}}

\begin{figure}[p]
  \centering
  \resizebox{!}{0.9\textheight}{\includegraphics{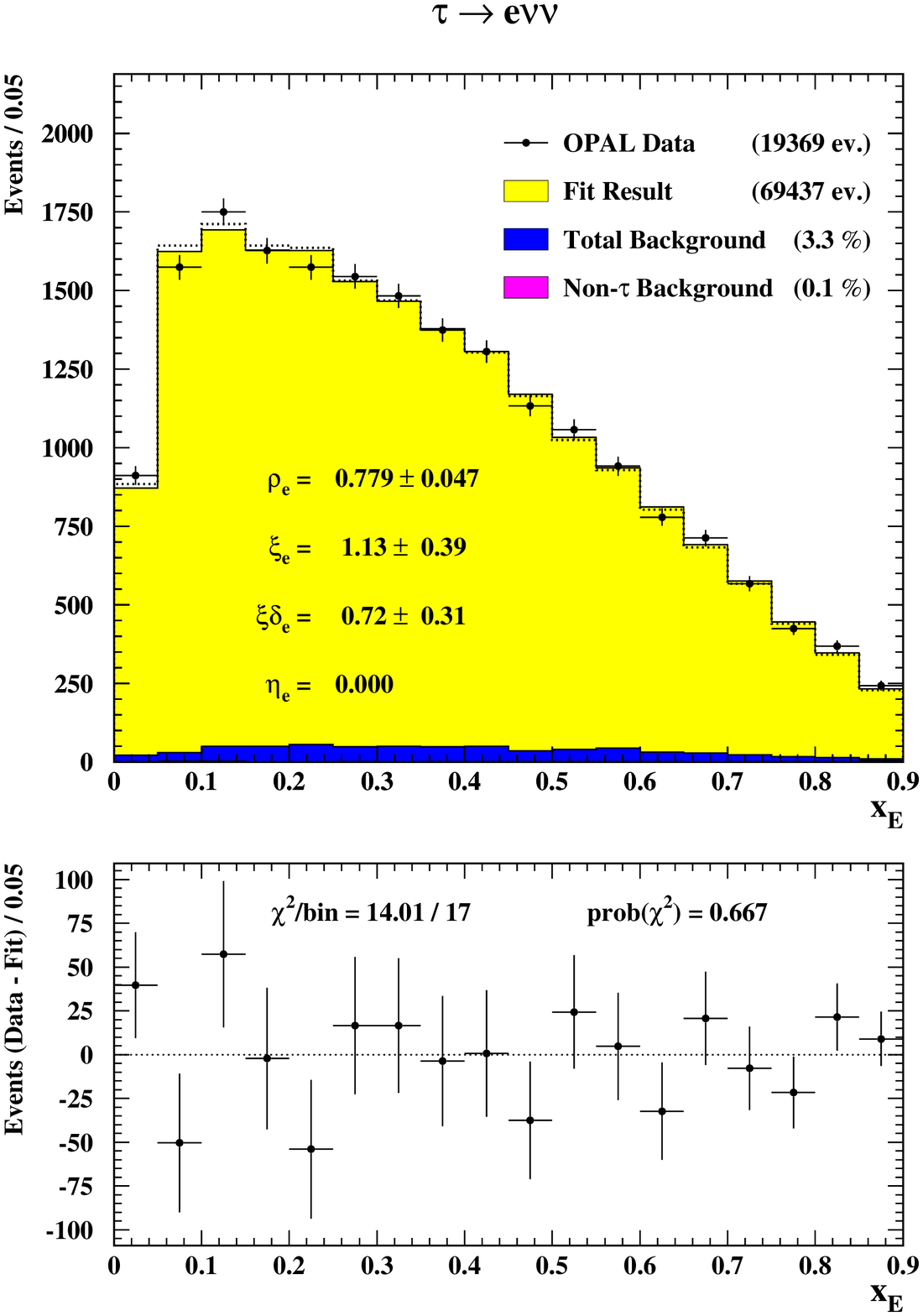}}
\caption{\em 
  The scaled \tautoe\ electron energy decay spectrum from \Pe--\Ph\ 
  events.  The quoted Michel parameters are the subset from the global
  fit that determines the plotted electron Monte Carlo spectrum. The
  dotted line represents the Standard Model expectation.}
\label{fig:single_e}
\end{figure}

\begin{figure}[p]
\centering
\resizebox{!}{0.9\textheight}{\includegraphics{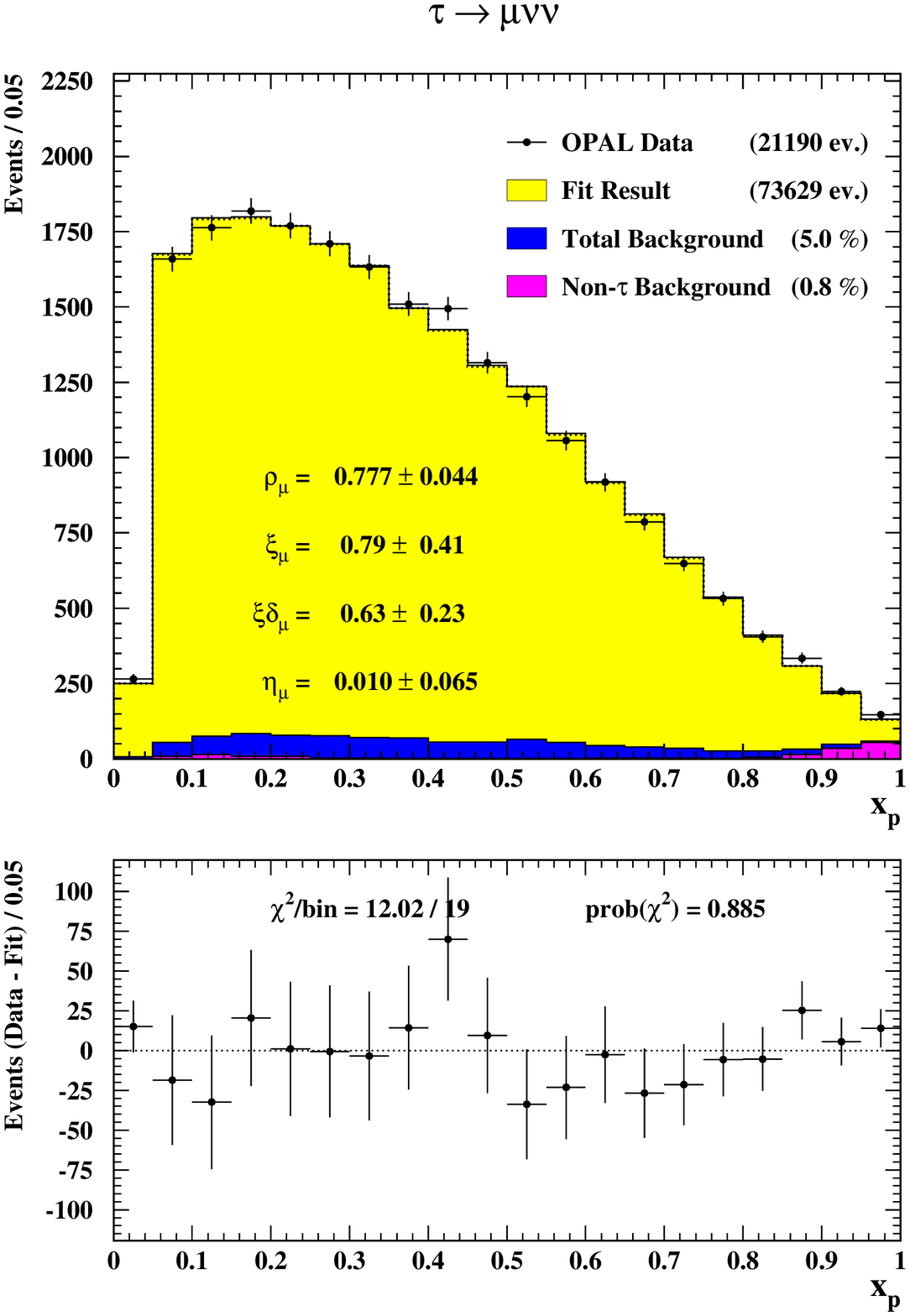}}
\caption{\em 
  The scaled \tautom\ muon momentum decay spectrum from \Pgm--\Ph\ 
  events.  The quoted Michel parameters are the subset from the global
  fit that determines the plotted muon Monte Carlo spectrum. The
  dotted line represents the Standard Model expectation.}
\label{fig:single_mu}
\end{figure}

The result of the most general global fit to single-\Pe, single-\Pgm\ 
and \Pe-\Pgm-correlations with seven free parameters is shown in
table~\ref{tab:result_8}. Table~\ref{tab:covariance_8} gives the
individual correlation coefficients.
\begin{table}[hbt]
\centering
\begin{tabular}{ccc@{\qquad}cc}
\\
\hline
\multicolumn{2}{c}{\tautoe} & \mspace &
\multicolumn{2}{c}{\tautom} \\
\hline
\rhoe & $\mbox{\RHOE} \pm \SRHOE  $  & & \rhom & $\mbox{\RHOM} \pm \SRHOM  $ \\
\xie  & $\mbox{\XIE } \pm \SXIE   $  & & \xim  & $\mbox{\XIM } \pm \SXIM   $ \\
\xde  & $\mbox{\XDE } \pm \SDELTAE$  & & \xdm  & $\mbox{\XDM } \pm \SDELTAM$ \\
\etae & 0 (fixed)                    & & \etam & $\mbox{\ETAM} \pm \SETAM  $ \\
\hline
\\
\end{tabular}
\caption{\em
Results of the global fit to the \Pe-\Ph, \Pgm-\Ph\ and \Pe-\Pgm\ energy
spectra. The first error reflects both data and Monte Carlo statistics, the
second error is systematic (see section~\ref{sec:systematic-errors}).
The error on $\etam$ includes the errors on the muonic branching ratio
and the \Pgt\ lifetime.
}

\label{tab:result_8}
\end{table}
\begin{table}[hbt]
\centering
\begin{tabular}{c|rrrrrr}
      & \mc\xie   & \mc\xde   & \mc\rhom   & \mc\xim   & \mc\xdm   & \mc\etam \\
\hline
\rhoe & \RHOEvXIE & \RHOEvXDE & \RHOEvRHOM & \RHOEvXIM & \RHOEvXDM & \RHOEvETAM \\
\xie  &           &  \XIEvXDE &  \XIEvRHOM &  \XIEvXIM &  \XIEvXDM &  \XIEvETAM \\
\xde  &           &           &  \XDEvRHOM &  \XDEvXIM &  \XDEvXDM &  \XDEvETAM \\
\rhom &           &           &            & \RHOMvXIM & \RHOMvXDM & \RHOMvETAM \\
\xim  &           &           &            &           &  \XIMvXDM &  \XIMvETAM \\
\xdm  &           &           &            &           &           &  \XDMvETAM \\
\end{tabular}
\caption{\em Correlation coefficients between the parameters of the global fit.}
\label{tab:covariance_8}
\end{table}
The first errors are due to data and Monte Carlo statistics as
obtained from the fit, the second ones are due to systematic
uncertainties (see section \ref{sec:systematic-errors}).
Figures \ref{fig:single_e} and \ref{fig:single_mu} show the single
decay spectra for \tautoe\ and \tautom\ decays, respectively. In both
figures, the light shaded histogram is the adjusted Monte Carlo
prediction using the Michel parameters from the global fit (table
\ref{tab:result_8}).

Note that the fit range for the \tautoe\ spectrum extends from $x = 0$
to $0.9$ while for \tautom\ as well as for the \Pe-\Pgm-correlation
spectra the full range ($x = 0$ to $1.0$) has been used. This is,
because the background from Bhabha events causes a relatively large
systematic uncertainty at high $x$ to single \tautoe\ decays which is
no longer present in the correlation spectra where the recoil is
required to be a \tautom\ decay.  For the \tautom\ single decay
(fig.~\ref{fig:single_mu}) the \eemm\ contamination at high $x$ is
well simulated and no restriction of the fit range is necessary.

In the plots below the single lepton spectra the differences between
data and Monte Carlo are shown.  Since only the shape of the
likelihood function is significant and not its scale, there is no
absolute criterion for the quality of the fit from the maximum
likelihood method itself.  As an estimator for the confidence level,
the $\chi^2$ probability\footnote{ It has to be noted that this
  $\chi^2$ probability is biased towards higher values. Since the
  number of fit parameters that constrain a specific spectrum cannot
  be assigned to the different event classes in a unique way the
  (number of bins - 1) has been taken as the number of degrees of
  freedom.}  for the plotted bins is appended to figures
\ref{fig:single_e} and \ref{fig:single_mu}.

\begin{figure}[p]
\centering
\resizebox{!}{0.9\textheight}{\includegraphics{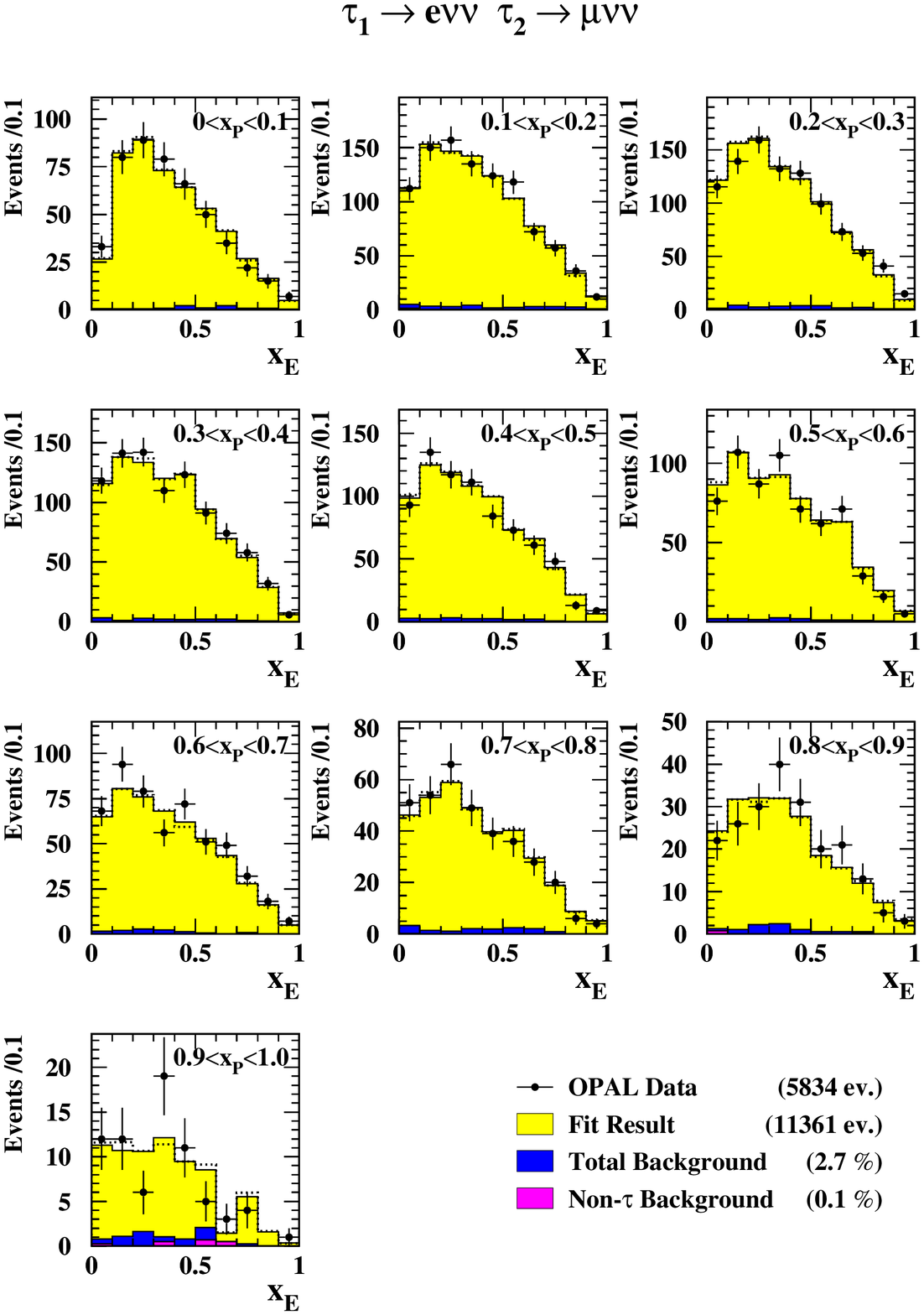}}
\caption{\em 
  The correlated \tautoe\ and \tautom\ decay spectra from \Pe--\Pgm\ 
  events together with the Monte Carlo distribution from the global
  fit. The spectrum of the scaled energy, $x_E$, of the decay electron
  is shown in slices for each bin of the scaled muon momentum, $x_p$.
  The dotted line represents the Standard Model expectation.}
\label{fig:double_e}
\end{figure}

\begin{figure}[p]
\centering
\resizebox{!}{0.9\textheight}{\includegraphics{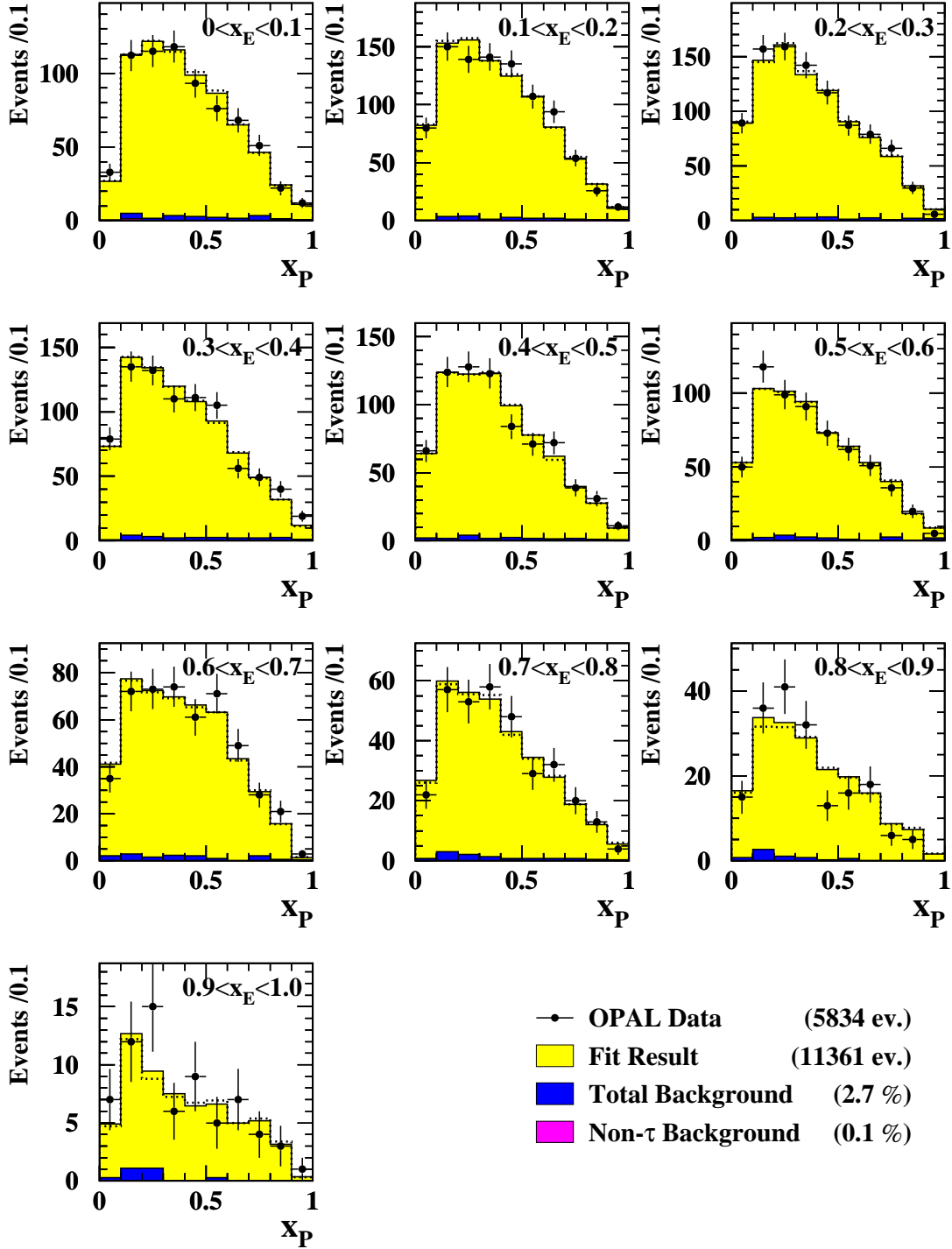}}
\caption{\em 
  The correlated \tautoe\ and \tautom\ decay spectra from \Pe--\Pgm\ 
  events together with the Monte Carlo distribution from the global
  fit. The spectrum of the scaled momentum, $x_p$, of the decay muon
  is shown in slices of each bin of the scaled electron energy, $x_E$.
  The dotted line represents the Standard Model expectation. (The
  plots show the same bins as figure~\ref{fig:double_e} in a different
  order.)}
\label{fig:double_mu}
\end{figure}

Figures \ref{fig:double_e} and \ref{fig:double_mu} are two different
representations of the \Pe-\Pgm-correlation spectrum, displaying
slices of $x_\Pe$ for fixed $x_\Pgm$ (fig.~\ref{fig:double_e}) and
vice versa (fig.~\ref{fig:double_mu}).  In all figures the dark shaded
histogram represents the total background from misidentified \Pgt\ 
decays and the black one the corresponding fraction of \eeee\ and
\eemm\ events. The background decomposition is listed in
table~\ref{tab:background}.
The fitted Michel parameters of table~\ref{tab:result_8} are
consistent with the V$-$A prediction of the Standard Model and
describe the spectra in figures \ref{fig:single_e}-\ref{fig:double_mu}
well.

\subsection{Systematic errors}
\label{sec:systematic-errors}

Various systematic uncertainties arising from different sources, both
detector specific and inherent to the method, have been investigated.
Besides these systematics, uncertainties concerning the Monte Carlo
simulation at the generator level, i.e., radiation effects, branching
fractions, process kinematics etc., have been found to be negligible.
The Monte Carlo statistical error is already reflected by the fit
errors as explained above.

For the study of systematic uncertainties imposed by the Monte Carlo
simulation various data control samples have been prepared. Two
samples, one of Bhabha and one of \Pgm-pair events, have been selected
using the criteria mentioned in section \ref{sec:tau-pair-selection}.
The energy and momentum distributions of these lepton-pair samples
show narrow peaks at the beam energy.
Two other samples with single lepton cones have been prepared using a
tagging technique. This has been done with preselected two-photon
events, \ggee\ and \ggmm, where one lepton is tagged based on a clear
decay mode identification so that the other cone can be investigated.
The same has been done with \eeee\ and \eemm\ events.

The following systematic effects, summarized in table
\ref{tab:systematic-errors}, remain:
\begin{itemize}

\item[-] The absolute scale of the energy and momentum measurement of
  electrons and muons, respectively, has been varied in the Monte
  Carlo.  The scale uncertainty has been determined by comparing
  \eeee\ and \eemm\ events from the data control samples with the
  corresponding Monte Carlo events. Rescaling the energies and momenta
  of the Monte Carlo events used in the fit within the observed
  uncertainty of $0.3\,\%$ and $0.2\,\%$, respectively, results in
  changes of the Michel parameters as given in
  table~\ref{tab:systematic-errors}.
  
\item[-] The uncertaintiy in the energy and momentum resolution has
  been determined from the same control samples which have been used
  for the previous study. Variation of the resolution leads to the
  quoted changes.
  
\item[-] The efficiency as determined from the Monte Carlo has been
  compared to the tagged control samples that include Bhabha and
  \Pgm-pair events covering the high $x$ region as well as two-photon
  events at low $x$. The events have been used by tagging one lepton
  while examining the efficiency for identifying the other lepton. The
  ratio of the efficiencies in data and Monte Carlo has been found to
  be consistent with unity.  The uncertainty on this efficiency ratio
  has been estimated by fitting the low and the high $x$ values with a
  straight line and determining the $1 \sigma$ bounds of a possible
  slope.  Since the overall efficiency is not used in the fit only an
  energy/momentum-dependent discrepancy could possibly affect the
  result.  This has been accounted for by weighting the Monte Carlo
  events within the uncertainty of the possible slopes.
  
\item[-] The background contributions from \Pgt\ and non-\Pgt\ sources
  which are estimated from the Monte Carlo simulation have been varied
  in the fit.  Previous studies of the decay mode identification
  mentioned in section \ref{sec:likelihood-identification} showed that
  the reference distributions of the likelihood variables agree with
  those of tagged data samples fairly well.  Reweighting the Monte
  Carlo such that it perfectly matches the data causes only slight
  changes in the resulting purities. The uncertainty of the background
  is savely estimated to be at the 10\,\% level, with the exception of
  the Bhabha background which has been varied by a factor of 2. The
  reason for this is a relatively poor modeling of the Bhabha
  background especially in the forward region.  Due to the small
  fraction of the background, the variation had no significant impact
  on the fit result.
  
\item[-] The dependence of the fit result on the energy and momentum
  range has been investigated by omitting the outer bins at low and
  high $x$ values. For the single \Pgm\ and the \Pe-\Pgm\ spectra the
  variation of the fit with and without the high $x$ bins also
  reflects the sensitivity to the description of the \eemm\ 
  background. No statistically significant change has been observed.
  
\item[-] Different sets of basis spectra have been used to study the
  influence of the particular choice of the Michel parameter basis on
  the description of the data. The same behavior has been observed as
  for the Monte Carlo study mentioned above with the cross checks of
  the fitting method. Due to the finite Monte Carlo statistics some
  bases were less sensitive to the Michel parameters than the
  preferred one, however, all alternative fits were consistent within
  their respective fit errors. This means that the basis choice does
  not contribute to the systematic error.

\end{itemize}

\begin{table}[hbt]
\centering
\begin{tabular}{lccc@{\qquad}cccc}

\hline
& $\Delta\rhoe$ & $\Delta\xie$ & $\Delta\xde$ &
  $\Delta\rhom$ & $\Delta\xim$ & $\Delta\xdm$ & $\Delta\etam$ \\
\hline
Energy scale                  & 0.017 & 0.11 & 0.11 &   $-$ &  $-$ &  $-$ &   $-$ \\
Momentum scale                &   $-$ &  $-$ &  $-$ & 0.013 & 0.04 & 0.04 & 0.001 \\
Energy resolution             & 0.003 & 0.03 & 0.04 &   $-$ &  $-$ &  $-$ &   $-$ \\
Momentum resolution           &   $-$ &  $-$ &  $-$ & 0.004 & 0.05 & 0.02 & 0.001 \\
Energy-dependent efficiency   & 0.023 & 0.08 & 0.08 &   $-$ &  $-$ &  $-$ &   $-$ \\
Momentum-dependent efficiency &   $-$ &  $-$ &  $-$ & 0.009 & 0.07 & 0.01 &   $-$ \\
\hline
Total                         & 0.029 & 0.14 & 0.14 & 0.016 & 0.09 & 0.05 & 0.001 \\
\end{tabular}
\caption{\em 
Contributions of systematic uncertainties to the error of each Michel
parameter for the global fit.  A hyphen indicates that the  listed effect
contributes less than 0.001 ($\rho,\eta$) or 0.01 ($\xi,\xi\delta$) to the
Michel parameter error. Note that the value of $\etam$ is dominated by the
branching ratio constraint.  }
\label{tab:systematic-errors}
\end{table}

\subsection{Lepton universality}

With the assumption of universality between electron and muon, i.e.,
with all couplings $g^\gamma_{\epsilon\omega}$ being the same for
\tautoe\ and \tautom, one set of Michel parameters can be used to
describe both leptonic decays. The fit then yields the results in
table~\ref{tab:result_4}. The systematic erros have been determined
using the same procedure as for the general fit.  The correlation
coefficients for this fit are listed in table~\ref{tab:covariance_4}.
The results are in agreement with the prediction of a V$-$A structure
of the charged weak current.
\begin{table}[hbt]
\centering
\begin{tabular}{cc}
\\
\hline
\multicolumn{2}{c}{\tautol} \mspace \\
\hline
\rhou & $\mbox{\RHOU} \pm \SRHOU  $ \\    
\xiu  & $\mbox{\XIU}  \pm \SXIU   $ \\
\xdu  & $\mbox{\XDU}  \pm \SDELTAU$ \\ 
\etau & $\mbox{\ETAU} \pm \SETAU  $ \\
\hline
\\
\end{tabular}
\caption{\em Result of the \Pe-\Pgm\ universality fit.}
\label{tab:result_4}
\end{table}
\begin{table}[hbt]
\centering
\begin{tabular}{c|rrr}
      & \mc\xiu   & \mc\xdu   & \mc\etau \\
\hline
\rhou & \RHOUvXIU & \RHOUvXDU & \RHOUvETAU \\
\xiu  &           &  \XIUvXDU &  \XIUvETAU \\
\xdu  &           &           &  \XDUvETAU \\
\end{tabular}
\caption{\em Correlation coefficients for the \Pe-\Pgm\ universality fit.}
\label{tab:covariance_4}
\end{table}

\subsection{Limits on the couplings}

\begin{figure}
\centering
\resizebox{!}{0.85\textheight}{\includegraphics{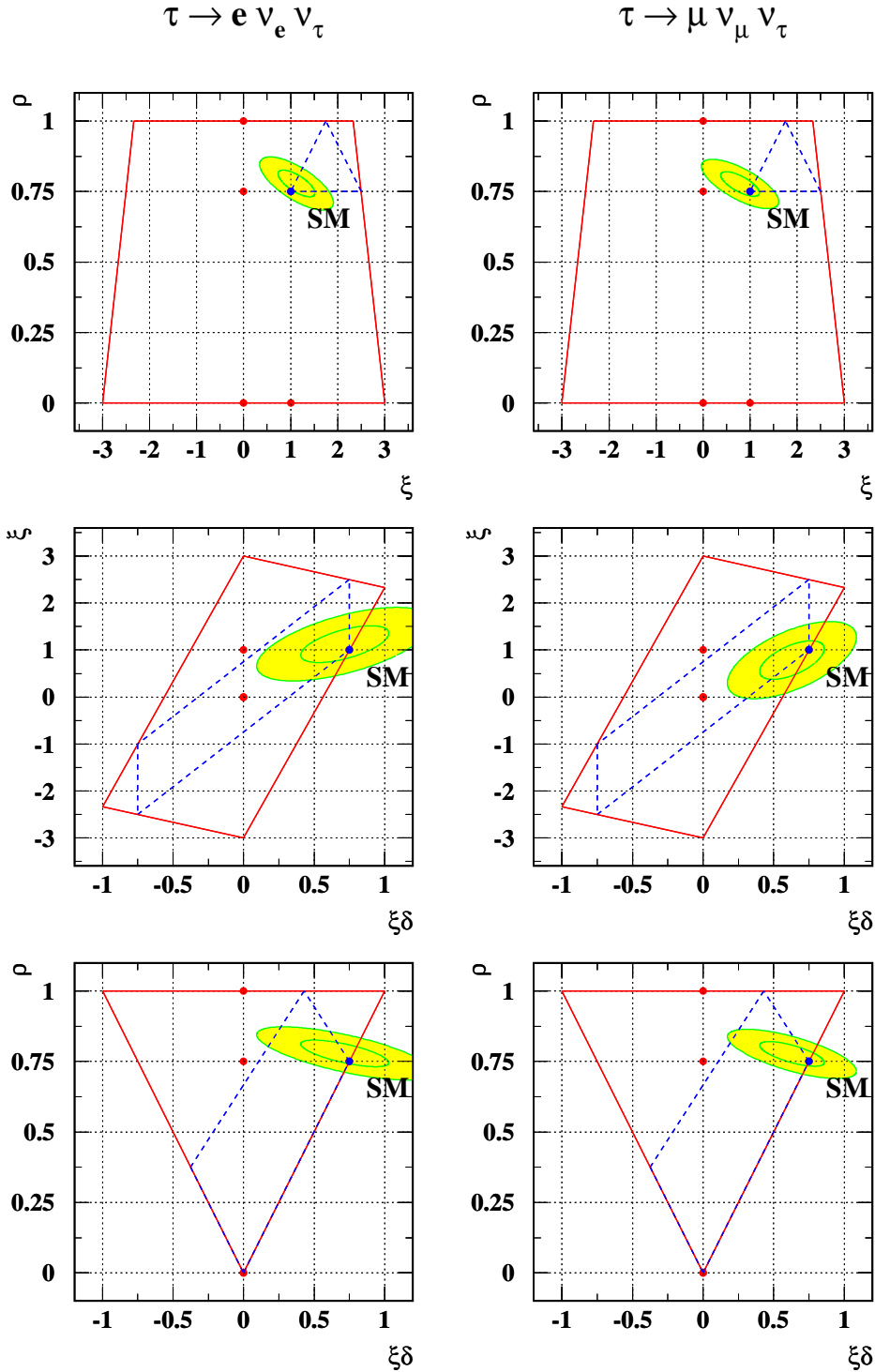}}
\caption{\em 
  Presentation of the global fit result in 2-dimensional projections
  of the Michel parameter space. Shown are the 1 and 2 $\sigma$
  ellipses of the fitted parameters.  The outer (solid) line encloses
  the physically allowed region of the parameters. The dashed line
  confines the area that remains when the parameter which is not
  plotted is fixed to its Standard Model value.  The closed circles
  indicate the basis spectra used in the fit. The labeled one
  represents the Standard Model.}
\label{fig:ellipses_8}
\end{figure}

From the measurement of the Michel parameters, limits on the absolute
values of the couplings $g_{\epsilon\omega}^{\gamma}$ (see
eq.~\ref{eqn:matrix-element}) can be extracted.
This is done by constructing positive-semidefinite expressions from
the measured parameters.
A general approach to find such expressions is to use the boundaries
of the physically allowed parameter space as shown in figure
\ref{fig:ellipses_8}.
In three dimensions, the physically allowed region of the three
parameters $\rho$, $\xi$ and $\xi\delta$ forms a tetrahedron
\cite{Rouge:tau-neutrino}.\footnote{The three plots in figure
  \ref{fig:ellipses_8} show the projections of this tetrahedron.}

An upper bound on $|\gvrl|$ as well as weak upper bounds on $|\gsrl|$
and $|\gtrl|$ can be set from the expression $1-\rho$.  Limits on
$|\gvrr|$ and $|\gsrr|$ can be retrieved from the expression $\rho -
\xi\delta$. Limits on the remaining couplings $|\gslr|$ and $|\gtlr|$
follow from the probability for the decay of a right-handed \Pgt\ 
lepton which is given below. An even stronger limit on $|\gvlr|$ can
be set by regarding a plane in the 3-dimensional parameter space
($\rho,\xi,\xi\delta$) which yields the expression $1 - \rho +
\frac{1}{3}\,\xi - \frac{7}{9}\,\xi\delta$.
The explicit dependence of these expressions on the couplings is:
\begin{eqnarray}
1-\rho
& = &
  \frac{1}{4} {|\gvll|}^2
+ {|\gvlr|}^2
+ {|\gvrl|}^2
+ \frac{1}{4} {|\gvrr|}^2
+ \frac{1}{16}{|\gsll|}^2
\nn \\
& &
+ \frac{1}{16}{|\gslr + 6\gtlr|}^2
+ \frac{1}{16}{|\gsrl + 6\gtrl|}^2
+ \frac{1}{16}{|\gsrr|}^2
, \\
\rho - \xd
& = &
  \frac{3}{2}{|\gvrr|}^2
+ \frac{3}{8}{|\gslr - 2\gtlr|}^2
+ \frac{3}{8}{|\gsrr|}^2
, \\
1 - \rho + \frac{1}{3}\,\xi - \frac{7}{9}\,\xi\delta
& = &
             {|\gvlr|}^2
+ \frac{1}{4}{|\gvrr|}^2
+ \frac{1}{16}{|\gslr + 6\gtlr|}^2
+ \frac{1}{16}{|\gsrr|}^2
.
\end{eqnarray}

Only the coupling $|\gsll|$ cannot be constrained since it cannot be
distinguished from the Standard Model coupling $|\gvll|$ on basis of
the four Michel parameters.
%
%
An upper bound on $|\gsll|$ would require the measurement of
correlations between one of the neutrinos and the charged
lepton~\cite{Jarlskog} or of the cross section for the process
\Pgngt\Pl\ra\Pgt\Pgnl~\cite{Fetscher:muon-decay}.

The probability that the \Pgt\ lepton decays as a right-handed particle
can be calculated as the sum of all couplings of the type
$g^{\gamma}_{\epsilon\mathrm{R}}$ normalized according to equation
\ref{eqn:normalization}:
%
\begin{eqnarray}
Q_\mathrm{R} & = &
    \frac{1}{4}{|\gslr|}^2 + \frac{1}{4}{|\gsrr|}^2
    + {|\gvlr|}^2 + {|\gvrr|}^2 + 3\,{|\gtlr|}^2
\nn \\
    & = &
    \frac{1}{2}\left( 1+ \frac{1}{3}\,\xi - \frac{16}{9}\,\xi\delta \right).
\end{eqnarray}

\begin{figure}
\centering
\resizebox{0.9\textwidth}{!}{\includegraphics{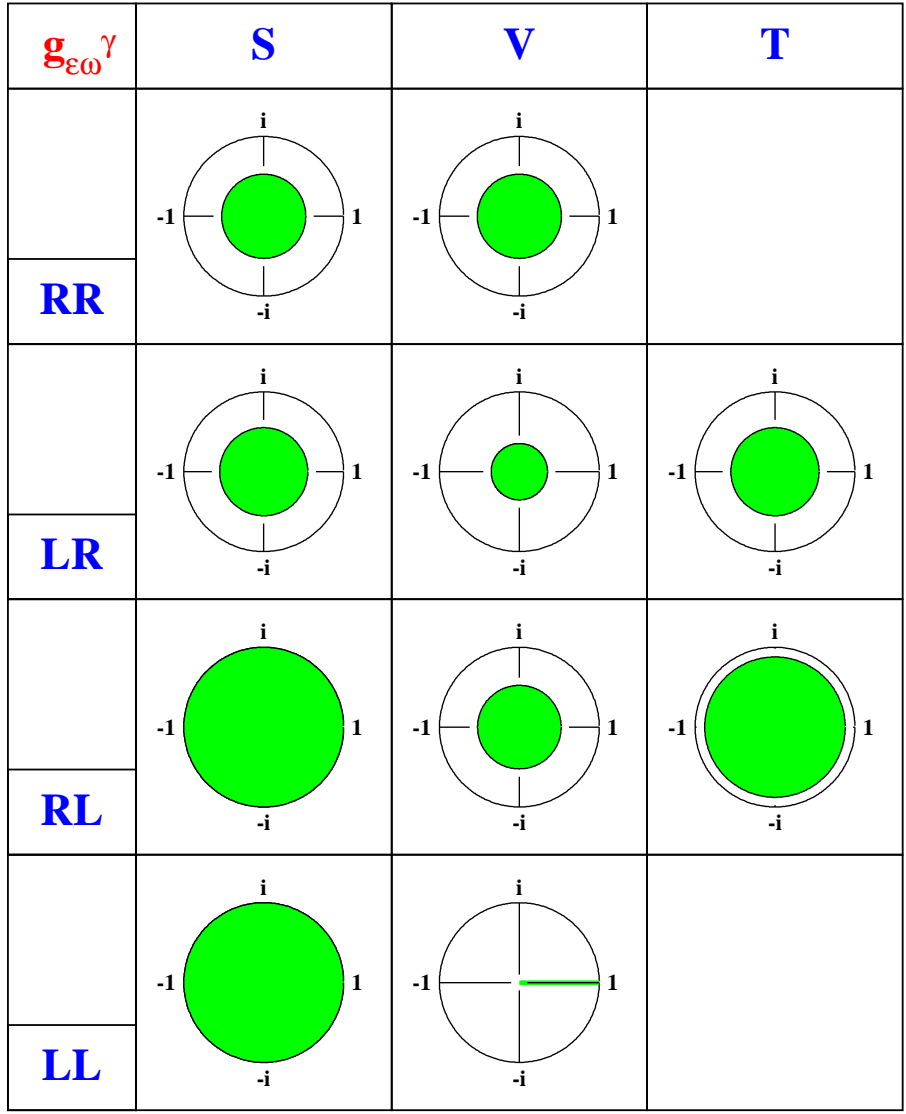}}
\caption{\em 
  90\,\% confidence limits on the normalized coupling constants
  $g_{\epsilon\omega}^{\gamma}/g^{\gamma}_\mathrm{max}$ (with
  $\gsmax=2$, $\gvmax=1$ and $\gtmax=\frac{1}{\sqrt{3}}$) under
  assumption of \Pe-\Pgm\ universality. The Standard Model coupling
  \gvll\ which is not constrained is chosen to be real and positive. }
\label{fig:limits_4}
\end{figure}

Using the correlations between the parameters the
\Pe-\Pgm\ universality fit yields:
\begin{equation}
   Q_\mathrm{R} = \mbox{\QtauRU}.
\end{equation}
From the result of the global and of the \Pe-\Pgm\ universality fit,
respectively, the bounds given in table~\ref{tab:limits} can be set at
the 90\,\% confidence level.
Figure~\ref{fig:limits_4} shows the limits on the universal coupling
constants normalized to their maximum values ($\gsmax=2$, $\gvmax=1$
and $\gtmax=\frac{1}{\sqrt{3}}$) introduced in
section~\ref{sec:ranges}.
\begin{table}[hbt]
\renewcommand{\arraystretch}{1.2}
\centering
\begin{tabular}{c|c|c|c}
          & \tautoe       & \tautom       & \tautol      \\
\hline
$|\gsrr|$ & $< \LIMgsrrE$ & $< \LIMgsrrM$ & $< \LIMgsrrU$ \\
$|\gslr|$ & $< \LIMgslrE$ & $< \LIMgslrM$ & $< \LIMgslrU$ \\
$|\gsrl|$ & $< \LIMgsrlE$ & $< \LIMgsrlM$ & $< \LIMgsrlU$ \\
$|\gsll|$ & $\leq 2     $ & $\leq 2     $ & $\leq 2     $ \\
\hline                                                    
$|\gvrr|$ & $< \LIMgvrrE$ & $< \LIMgvrrM$ & $< \LIMgvrrU$ \\
$|\gvlr|$ & $< \LIMgvlrE$ & $< \LIMgvlrM$ & $< \LIMgvlrU$ \\
$|\gvrl|$ & $< \LIMgvrlE$ & $< \LIMgvrlM$ & $< \LIMgvrlU$ \\
$|\gvll|$ & $\leq 1     $ & $\leq 1     $ & $\leq 1     $ \\
\hline                                                    
$|\gtlr|$ & $< \LIMgtlrE$ & $< \LIMgtlrM$ & $< \LIMgtlrU$ \\
$|\gtrl|$ & $< \LIMgtrlE$ & $< \LIMgtrlM$ & $< \LIMgtrlU$ \\
\end{tabular}
\caption{\em 
  90\,\% confidence limits on the coupling constants for the global fit
  and under the assumption of \Pe-\Pgm\ universality. As mentioned in the
  text, no limits can be set on the couplings \gsll\ and \gvll\ which
  are listed only for completeness.}
\label{tab:limits}
\end{table}

\subsection{Mass of a charged Higgs boson}

In models with two scalar field doublets, such as the Minimal
Supersymmetric Standard Model (MSSM), the existence of a charged Higgs
boson is assumed which contributes to the \Pgt\ decay through a scalar
coupling. The value of the additional coupling is, assuming vanishing
neutrino masses
\cite{Haber:Higgs-tau-charm,Hollik:Higgs,McWilliams:Higgs}:
\begin{equation}
\gs_\Pl = - m_\Pl\ m_\Pgt \left(\frac{\tan\beta}{m_{\PHpm}}\right)^2.
\end{equation}
Here $m_{\PHpm}$ is the mass of the charged Higgs boson, $\tan\beta$
is the ratio of the vacuum expectation values of the two Higgs
doublets and \Pl\ denotes either \Pe\ or \Pgm. Under the assumption
that the neutrinos are still left-handed, the couplings are of the
type \gsrr. After applying the normalization $N_\Pl = |\gvll|^2 +
\frac{1}{4} |\gsrrl|^2$, the Michel parameters can be written
as:\footnote{This can be verified by inserting the normalized \gvll\ 
  and \gsrr\ into the definition of the Michel parameters given in
  equation~\ref{eqn:michel-parameters}.}
\begin{eqnarray}
\rhol & = & \frac{3}{4}
\nn \,, \\
\xil  & = & \frac{1 - (\gsrrl/2)^2}{1 + (\gsrrl/2)^2}
\,, \\
\xdl  & = & \frac{3}{4}\,\xil
\nn \,, \\
\nn \etal & = & - \frac{\gsrrl/2}{1 + (\gsrrl/2)^2}
%
\,.
\end{eqnarray}

Using the above relations, the value of $m_{\PHpm}/\tan\beta$ can be
fitted directly to the data.
The likelihood function saturates for high Higgs boson masses or small
values of $\tan\beta$.  From the log likelihood a limit can be
extracted as:\footnote{The following limits are determined as the
  value at which the log likelihood has dropped by the amount that
  corresponds to the quoted confidence level 
  .}
\begin{equation}
 \mbox{\LIMHiggs}.
\end{equation}

\subsection{Left-right symmetric model}

In left-right symmetric models, parity violation of the charged
current is caused by spontaneous symmetry breaking. In such models a
second \PW\ boson is assumed~\cite{Polak,Polak:update}. The mass
eigenstates $\PW_{1,2}$ are not necessarily identical to the weak
eigenstates $\PW_\mathrm{L,R}$ as mixing can occur.  The model is
parameterized by the mass ratio $\beta$ of the physical eigenstates,
\begin{equation}
\beta = \frac{m_\PWi^2}{m_\PWii^2},
\end{equation}
and by the mixing angle $\zeta$ that connects the physical masses to
the masses of the weak eigenstates:
\begin{equation}
m_\mathrm{W_{1,2}}^2 = \frac{1}{2} \left(
m_\PWL^2 + m_\PWR^2 \pm \frac{m_\PWL^2 - m_\PWR^2}{\cos 2\zeta} \right).
\end{equation}
In this model the Michel parameters can be written as:
\begin{eqnarray}
\rho      & = & \frac{3}{4} \cos^4\zeta \left( 1 + \tan^4\zeta + 
                \frac{4\beta}{1+\beta^2}\tan^2\zeta \right)
\nn \,, \\ 
\xi       & = & \cos^2\zeta \left(1-\tan^2\zeta \right) 
                \frac{1-\beta^2}{1+\beta^2}
\,, \\
\xi\delta & = & \frac{3}{4}\,\xi
\nn \,, \\
\eta      & = & 0
\nn \,.
\end{eqnarray}
A limit on $\beta$ can be transformed into a limit on $m_\PWii$ by
using the direct measurement of the \PW\ mass: $m_\PWi =
\MW$~\cite{LEP:EW}.

\begin{figure}
  \centering
  \resizebox{0.9\textwidth}{!}{\includegraphics{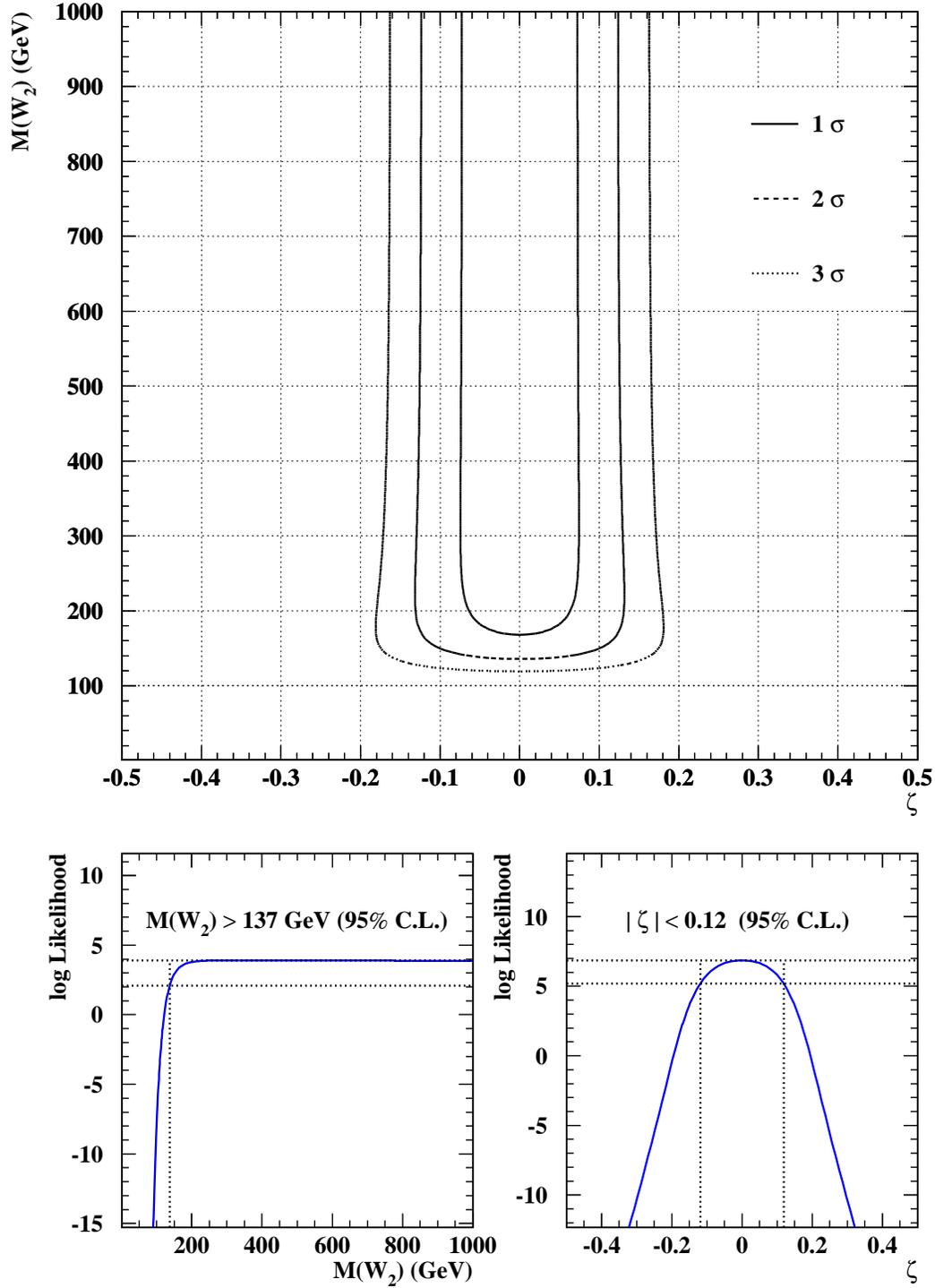}}
\caption{\em
  Log likelihood as a function of the mass of a second \PW\ boson,
  $m_\PWii$, and the mixing angle $\zeta$ in the left-right symmetric
  model (upper plot). From this likelihood function limits on
  $m_\PWii$ (lower left plot) and $\zeta$ (lower right plot) are
  calculated after integration of the likelihood over the second
  variable.}
\label{fig:mrightw_proj}
\end{figure}

For arbitrary mixing the upper plot of figure~\ref{fig:mrightw_proj}
shows the one, two and three $\sigma$ contours of the log likelihood
as a function of $\beta$ and $\zeta$.
Integration of the two-dimensional likelihood over $\zeta$ yields the
corresponding function shown in in the lower left plot of the same
figure.  From this likelihood a limit on $m_\PWii$ which is valid for
arbitrary mixing can be extracted as:
\begin{equation}
 \mbox{\LIMWii}.
\end{equation}
Similarly, integration over $m_\PWii$ allows one to set bounds on the
mixing angle independently of the \PWii\ mass (lower right plot):
\begin{equation}
 \mbox{\LIMzeta}.
\end{equation}

For $\zeta=0$, $\PWii$ and $\PWR$ become identical, and a limit on
$m_\PWR$ can be given from a fit of the Michel parameter $\xi$ alone.
In this case there is no mixing but an additional coupling to a pure
right-handed \PW\ that is proportional to its inverse mass:\footnote{
  Inserting the normalized \gvll\ and \gvrr\ into the definition of
  the Michel parameters (eq.~\ref{eqn:michel-parameters}) yields the
  above relations for $\zeta=0$.}
\begin{equation}
  \gv_\mathrm{LL,RR} \sim \frac{1}{m_{\PW_\mathrm{L,R}}}.
\end{equation}
Under the assumption of no mixing the following limit is extracted
from the log likelihood function:
\begin{equation}
 \mbox{\LIMWR}.
\end{equation}

\section{Summary}

The Michel parameters of the leptonic \Pgt\ decays have been measured
from the data collected with the OPAL detector in the years 1990 to
1995. The parameters \rhol, \xil, \xdl\ and \etam\ (with $\Pl =
\Pe,\Pgm$) were extracted from the energy spectra of the charged decay
leptons and from their energy-energy correlations. A new method has
been presented which involves a global likelihood fit of Monte Carlo
generated events with radiative corrections at the generator level and
complete detector simulation and background treatment. In the
framework of the most general Lorentz structure for both leptonic
decays the result of the global fit is:
\begin{center}
\begin{tabular}{r@{$\ \ =\ \ $}l@{\qquad}r@{$\ \ =\ \ $}l}
\rhoe & $\mbox{\RHOE}\pm\SRHOE  $, &
\rhom & $\mbox{\RHOM}\pm\SRHOM  $, \\
 \xie & $\mbox{\XIE}\pm\SXIE    $, &
 \xim & $\mbox{\XIM}\pm\SXIM    $, \\
 \xde & $\mbox{\XDE}\pm\SDELTAE $, &
 \xdm & $\mbox{\XDM}\pm\SDELTAM $, \\
 \multicolumn{2}{l}{}              &
\etam & $\mbox{\ETAM}\pm\SETAM  $, \\
\end{tabular}
\end{center}
where the value of \etau\ is dominated by a constraint using the
previously published values of the leptonic branching ratios and the
\Pgt\ lifetime. The \Pgt\ polarization has been inferred from neutral
current data.  The Michel parameters have also been measured under the
assumption of \Pe-\Pgm\ universality and in terms of specific models.
The \Pe-\Pgm\ universilty fit yields:
\begin{center}
  \begin{tabular}{r@{$\ \ =\ \ $}l@{\qquad}r@{$\ \ =\ \ $}l}
    \rhou & $\mbox{\RHOU} \pm \SRHOU   $, &
    \xiu  & $\mbox{\XIU}  \pm \SXIU    $, \\
    \etau & $\mbox{\ETAU} \pm \SETAU   $, &
    \xdu  & $\mbox{\XDU}  \pm \SDELTAU $.
  \end{tabular}
\end{center}
Limits have been obtained on the individual coupling constants as well
as on the masses of new gauge bosons, such as a right-handed \PW\ 
boson, \PWR, and a charged Higgs boson.
No indication for new physics processes has been observed.
The results are in agreement with the V$-$A prediction of the Standard
Model.

\newpage


\begin{table}[p]
\centering
\renewcommand{\arraystretch}{1.6}
\scriptsize
\begin{tabular}{|c|c|cccc|}
\hline
Type of interaction & Coupling constants &
\multicolumn{4}{|c|}{Michel parameters} \\
\Pl-\Pgnl-vertex \ot\ \Pgt-\Pgngt-vertex & &
$\rho$ & $\xi$ & $\xi\delta$ & $\eta$ \\
\hline\hline
V$-$A \ot\ V$-$A & $\gvll =1$
               & $3/4$ & $\m1$ & $\m3/4$ & $0$ \\
V$+$A \ot\ V$+$A   & $\gvrr =1$
               & $3/4$ &  $-1$ & $-3/4$ & $0$ \\
V \ot\ V       & $\gvll =\gvrl =\gvlr =\gvrr =1/2$
               & $3/8$ & $\m0$ & $\m0$ & $0$ \\
A \ot\ A       & $\gvll =-\gvrl =-\gvlr =\gvrr =1/2$
               & $3/8$ & $\m0$ & $\m0$ & $0$ \\
V$-$A \ot\ V$+$A  & $\gvlr =1$
               & $0$ & $\m3$ & $\m0$ & $0$ \\
V$+$A \ot\ V$-$A  & $\gvrl =1$
               & $0$ &  $-3$ & $\m0$ & $0$ \\
\hline
S$+$P \ot\ S$-$P  & $\gsll =2$
               & $3/4$ & $\m1$ & $\m3/4$ & $0$ \\
S$-$P \ot\ S$+$P  & $\gsrr =2$
               & $3/4$ & $-1$ & $-3/4$ & $0$ \\
S \ot\ S       & $\gsll =\gsrl =\gslr =\gsrr =1$
               & $3/4$ & $\m0$ & $\m0$ & $0$ \\
P \ot\ P       & $-\gsll =\gsrl =\gslr =-\gsrr =1$
               & $3/4$ & $\m0$ & $\m0$ & $0$ \\
S$+$P \ot\ S$+$P   & $\gslr =2$
               & $3/4$ & $-1$ & $-3/4$ & $0$ \\
S$-$P \ot\ S$-$P & $\gsrl =2$
               & $3/4$ & $\m1$ & $\m3/4$ & $0$ \\
\hline
T \ot\ T       & $\gtlr =\gtrl =\sqrt{1/6}$
               & $1/4$ & $\m0$ & $\m0$ & $0$ \\
50\% V$-$A \ot\ V$-$A, 50\% S \ot\ S    & $\gvll =\gs =\sqrt{1/2}$
               & $3/4$ & $\m1/2$ & $\m3/8$ & $1/4$ \\
50\% V$-$A \ot\ V$-$A, 50\% S$-$P \ot\ S$+$P
               & $\gvll =\sqrt{1/2},\; \gsrr =\sqrt{2}$
               & $3/4$ & $\m1$ & $\m0$ & $1/2$ \\
50\% V$\mp$A \ot\ V$\mp$A, 50\% S$\pm$P \ot\ S$\mp$P
               & $\gvll = \gvrr = 1/2,\; \gsrr = \gsll =1$
               & $3/4$ & $\m0$ & $\m0$ & $1/2$ \\
50\% V$-$A \ot\ V$-$A, 50\% V$+$A \ot\ V$-$A & $\gvll =\gvrl =\sqrt{1/2}$
               & $3/8$ & $-1$ & $\m3/8$ & $0$ \\
50\% V$+$A \ot\ V$+$A, 50\% V$-$A \ot\ V$+$A & $\gvrr =\gvlr =\sqrt{1/2}$
               & $3/8$ & $\m1$ & $-3/8$ & $0$ \\
50\% V$-$A \ot\ V$+$A, 50\% V$+$A \ot\ V$-$A & $\gvlr =\gvrl =\sqrt{1/2}$
               & $0$ & $\m0$ & $\m0$ & $0$ \\
67\% V$-$A \ot\ V$+$A, 33\% V$+$A \ot\ V$-$A
               & $\gvlr =\sqrt{2/3},\; \gvrl =\sqrt{1/3}$
               & $0$ & $\m1$ & $\m0$ & $0$ \\
75\% S$\pm$P \ot\ S$\pm$P, 25\% T \ot\ T
               & $\gslr =\gsrl =\sqrt{3/2},\; \gt =-\sqrt{1/24}$
               & $1$ & $\m0$ & $\m0$ & $0$ \\
12.5\% S$\pm$P \ot\ S$\pm$P, 50\% V$\mp$A \ot\ V$\pm$A, 37.5\% T \ot\ T
               & $\gslr =\gsrl =\gvlr =\gvrl =1/2,\; \gt =1/4$
               & $0$ & $\m0$ & $\m0$ & $1$ \\
\hline
\end{tabular}
\caption{\em Example coupling constants and corresponding Michel parameters}
\label{tab:couplings}
\end{table}

\medskip
\bigskip\bigskip\bigskip
\appendix
\par
Acknowledgements:
\par
We particularly wish to thank the SL Division for the efficient operation
of the LEP accelerator at all energies
 and for their continuing close cooperation with
our experimental group.  We thank our colleagues from CEA, DAPNIA/SPP,
CE-Saclay for their efforts over the years on the time-of-flight and trigger
systems which we continue to use.  In addition to the support staff at our own
institutions we are pleased to acknowledge the  \\
Department of Energy, USA, \\
National Science Foundation, USA, \\
Particle Physics and Astronomy Research Council, UK, \\
Natural Sciences and Engineering Research Council, Canada, \\
Israel Science Foundation, administered by the Israel
Academy of Science and Humanities, \\
Minerva Gesellschaft, \\
Benoziyo Center for High Energy Physics,\\
Japanese Ministry of Education, Science and Culture (the
Monbusho) and a grant under the Monbusho International
Science Research Program,\\
German Israeli Bi-national Science Foundation (GIF), \\
Bundesministerium f{\"u}r Bildung, Wissenschaft,
Forschung und Technologie, Germany, \\
National Research Council of Canada, \\
Research Corporation, USA,\\
Hungarian Foundation for Scientific Research, OTKA T-016660, 
T023793 and OTKA F-023259.\\

\clearpage
\bibliography{pr245}

\end{document}